\documentclass[12pt,english,floatfix,nofootinbib,superscriptaddress,aps,prd,preprint]{revtex4}
\usepackage[utf8]{inputenc}
\usepackage{float}
\usepackage{array}
\usepackage{lipsum}
\usepackage{dsfont}
\usepackage{graphicx}
\usepackage{amsmath,amsthm,amsfonts,amssymb}
\usepackage{graphicx}
\usepackage[english]{babel} 
\usepackage{color}
\usepackage{tensor}
\usepackage{esint}
\usepackage[dvips]{epsfig}
\usepackage[dvips]{graphicx}
\usepackage{float}
\usepackage{units}
\usepackage{textcomp}
\usepackage{mathrsfs}
\usepackage{amsmath}
\usepackage[makeroom]{cancel}
\usepackage{amssymb}
\usepackage{amsbsy}
\usepackage{amsfonts}
\usepackage{amssymb,mathrsfs,xcolor}
\usepackage{esint}
\usepackage{braket}
\usepackage{array}
\usepackage{graphicx}

\usepackage{wasysym}
\usepackage{multirow}
\usepackage{wrapfig}
\usepackage{subfig}

\usepackage{stmaryrd}
\usepackage{upgreek}

\makeatletter

\makeatletter\usepackage{babel}

\usepackage{hyperref}
\hypersetup{
    colorlinks,
    citecolor=blue,
    filecolor=green,
    linkcolor=purple,
    urlcolor=red,
}

\usepackage{slashed}

\newcommand{\ie}{\begin{equation}}
\newcommand{\fe}{\end{equation}}
\newcommand{\se}{\begin{eqnarray}}
\newcommand{\ff}{\end{eqnarray}}

\begin{document}

\title{Fermions on a torus knot}

\author{A. A. Ara\'{u}jo Filho}
\email{dilto@fisica.ufc.br}

\affiliation{Universidade Federal do Cear\'a (UFC), Departamento de F\'isica,\\ Campus do Pici,
Fortaleza -- CE, C.P. 6030, 60455-760 -- Brazil.}

\author{J. A. A. S. Reis}
\email{jalfieres@gmail.com}

\affiliation{Universidade Federal do Maranh\~{a}o (UFMA), Departamento de F\'{\i}sica,\\ Campus Universit\'{a}rio do Bacanga, S\~{a}o Lu\'{\i}s -- MA, 65080-805, -- Brazil}

\affiliation{Universidade Estadual do Maranh\~{a}o (UEMA), Departamento de F\'{i}sica,\\ Cidade Universit\'{a}ria Paulo VI, S\~{a}o Lu\'{i}s -- MA, 65055-310, -- Brazil.}

\author{Subir Ghosh}
\email{subirghosh20@gmail.com}

\affiliation{Physics and Applied Mathematics Unit, Indian Statistical Institute, 203 B.T.Road, Kolkata 700108, India.}


\date{\today}

\begin{abstract}

In this work, we investigate the effects of a nontrivial topology (and geometry) of a system considering \textit{interacting} and \textit{noninteracting} particle modes, which are restricted to follow a closed path over the torus surface. In order to present a prominent thermodynamical investigation of this system configuration, we carry out a detailed analysis using statistical mechanics within the grand canonical ensemble approach to deal with \textit{noninteracting} fermions. In an analytical manner, we study the following thermodynamic functions in such context: the Helmholtz free energy, the mean energy, the magnetization and the susceptibility. Further, we take into account the behavior of Fermi energy of the thermodynamic system. Finally, we briefly outline how to proceed in case of \textit{interacting} fermions. 

\end{abstract}

\maketitle


\section{Introduction}\label{Sec:Properties}

The knowledge of thermodynamic properties of  many body systems plays an essential role in condensed matter physics, especially in case of developing new materials \cite{gaskell2012introduction,muhlschlegel1972thermodynamic,dehoff2006thermodynamics,jones1974thermodynamics,davidovits1991geopolymers,safran2018statistical,bates1990block}. Electrons in metals, under widely varying circumstances, behave as free particles and can be treated as an electron gas under certain approximations \cite{araujo2017structural,silva2018polarized,casida1998molecular,lee1988development,segall2002first,perdew1981self,perdew1986density,perdew1992accurate,perdew1996generalized}. In this regard, surface effects such as shape dependence \cite{dai2003quantum,dai2004geometry,potempa1998dependence} and nontrivial topology  \cite{angelani2005topology,bendsoe2013topology,oliveira2020relativistic,narang2021topology} can lead to some novel and unexpected features. A generic computational problem is to derive the partition function that requires a sum over all accessible quantum states \cite{landau2013course,reichl1999modern,huang2009introduction}. For higher temperatures, the particle effective wavelength turns out to be too short in comparison with the characteristic dimension of the system and, therefore, the boundary effects can be overlooked. Previously, such idea was exploited by Rayleigh and Jeans in the radiation theory of electromagnetism \cite{zettili2003quantum,purcell1965electricity,tipler2003modern}; moreover, this was corroborated in a purely mathematical framework by Weyl \cite{weyl1968gesammelte}.

Quantum mechanics of particle on a torus knot was analyzed in Ref. \cite{sreedhar2015classical}. Further topological aspects such as Berry phase and Hannay angle features to this model were investigated in \cite{ghosh2018particle,das2016particle,biswas2020quantum}; in particular, curvature and torsion effects of the knotted path were also taken into account. The generic problem of a quantum particle restricted to a curved path or to a general surface, was considered earlier in \cite{da1981quantum,wang2018geometric,ortix2015quantum,da2017quantum,ferrari2008schrodinger}.

Knot theory is a well known branch of pure mathematics \cite{tu2011manifolds,tu2017differential}. The quantum mechanics on particles on torus knot can be motivated by the carbon nanorings (made from carbon nanotubes \cite{liu2001structural}) that are intimately connected to the edges of nanotubes, nanochains, graphene, and nanocones \cite{kharissova2019inorganic, chen2005mechanical}. Such nanorings (or nanotori) have a variety of notable features, being mostly investigated in generic calculations \cite{madani2012energies} and  specializing  in multiresonant properties \cite{lewicka2013nanorings}, magneto-optical activity \cite{feng2014magnetoplasmonic}, paramagnetism \cite{liu2002colossal}, and ferromagnetism \cite{vojkovic2016magnetization, rodriguez2004magnetism}. Moreover, these present nanostructures may also be employed in optical communications \cite{du2014silicon}, isolators, traps for ions and atoms \cite{chan2012carbon}, and lubricants \cite{pena2019novel}.

In this sense, the present work is aimed at studying the topological effects of a novel thermodynamic system comprising a gas of fermions, that follows a closed path with nontrivial knot structure on the surface of a torus. For convenience, one might visualize the particles moving within a carbon nanotube wound around the torus in the form of a knot. Furthermore, we calculate the Helmholtz free energy,  mean energy, magnetization and susceptibility of the system. In our investigation, we exploit the formalism of \cite{araujo2021thermal,reis2020does} to derive the thermodynamic quantities -- using  results from \cite{das2016particle}, for both \textit{interacting} and \textit{noninteracting} fermion particles.

The paper is organized as follows: in Section II, the basic formalism and notations are set up by considering a system of noninteracting particles and analytic expressions of all the thermodynamic quantities are computed starting from the partition function. Noninteracting fermions were introduced in Section III and, together with other observables, the Fermi energy is also studied. In Section IV, we outline the way to incorporate interactions in the fermionic model. The paper is concluded with a summary of the results obtained in Section V.   
\section{Statistical mechanics  of $N$ particles  in a torus knot subjected to a magnetic field}\label{Sec:N-torus}
Let us consider an ensemble composed of $N$ particles moving on a single torus in a prescribed path. For the situation that we are considering, the canonical ensemble theory is sufficient to describe this case. The partition function for the system of $N$ particle is given by%
\begin{equation}
\mathcal{Z}=\sum_{\left\{ \Omega \right\} }\exp \left( -\beta E_{\Omega
}\right) ,  \label{eq:Partition-function}
\end{equation}%
where $\Omega $ is related to all accessible quantum states. Since we are
dealing with noninteracting particles, the partition function $\left( \ref%
{eq:Partition-function}\right) $ can be factorized, giving the following result
below%
\begin{equation}
\mathcal{Z}=\mathcal{Z}_{1}^{N}=\left\{ \sum_{\left\{ \Omega \right\} }\exp
\left( -\beta E_{\Omega }\right) \right\} ^{N},
\label{eq:Partition-function-1}
\end{equation}%
where we have defined the single partition function as
\begin{equation}
\mathcal{Z}_{1}=\sum_{\Omega }\exp \left( -\beta E_{\Omega }\right) .
\label{eq:Single-Partition-function}
\end{equation}%
It is interesting to point out that, even though the underling geometry is not a simply connected one, there is no problem in regarding the factorization above, since there is no interaction potential generated by the geometry itself. This affirmation can naturally be checked by looking at the energy spectrum given by Eq. \ref{q:SEnergy}\footnote{Note that the procedure used here to carried out the factorization is not a general result. It suits only for the results presented in this paper.}. The connection with macroscopic thermodynamic observables can be given by  Helmholtz free energy%
\begin{equation}
f=-\frac{1}{\beta }\lim_{N\rightarrow \infty }\frac{1}{N}\ln \mathcal{Z},
\label{eq:Helmm-f}
\end{equation}%
where $f$ stands for the  Helmholtz free energy per particle. With this quantity, we can derive the following
thermodynamic state functions%
\begin{equation}
s=-\frac{\partial f}{\partial T},\phantom{ss}c=T\frac{\partial s}{\partial T},\phantom{ss}u=-\frac{\partial }{\partial \beta }\ln \mathcal{Z},
\label{eq:Therm-functions1}
\end{equation}%
where $s,c,u$ represent entropy, heat capacity and energy
per particle, respectively. We can also calculate  magnetization $\mathfrak{m}$ and  susceptibility $\chi $ with respect to the external magnetic $B$, as given by
\begin{equation}
\mathfrak{m}=-\frac{\partial f}{\partial B},\phantom{ss}\chi=\frac{\partial \mathfrak{m}}{\partial B}.
\label{eq:Therm-functions2}
\end{equation}%

\begin{figure}[h!]
\centering
\includegraphics[scale=1.0]{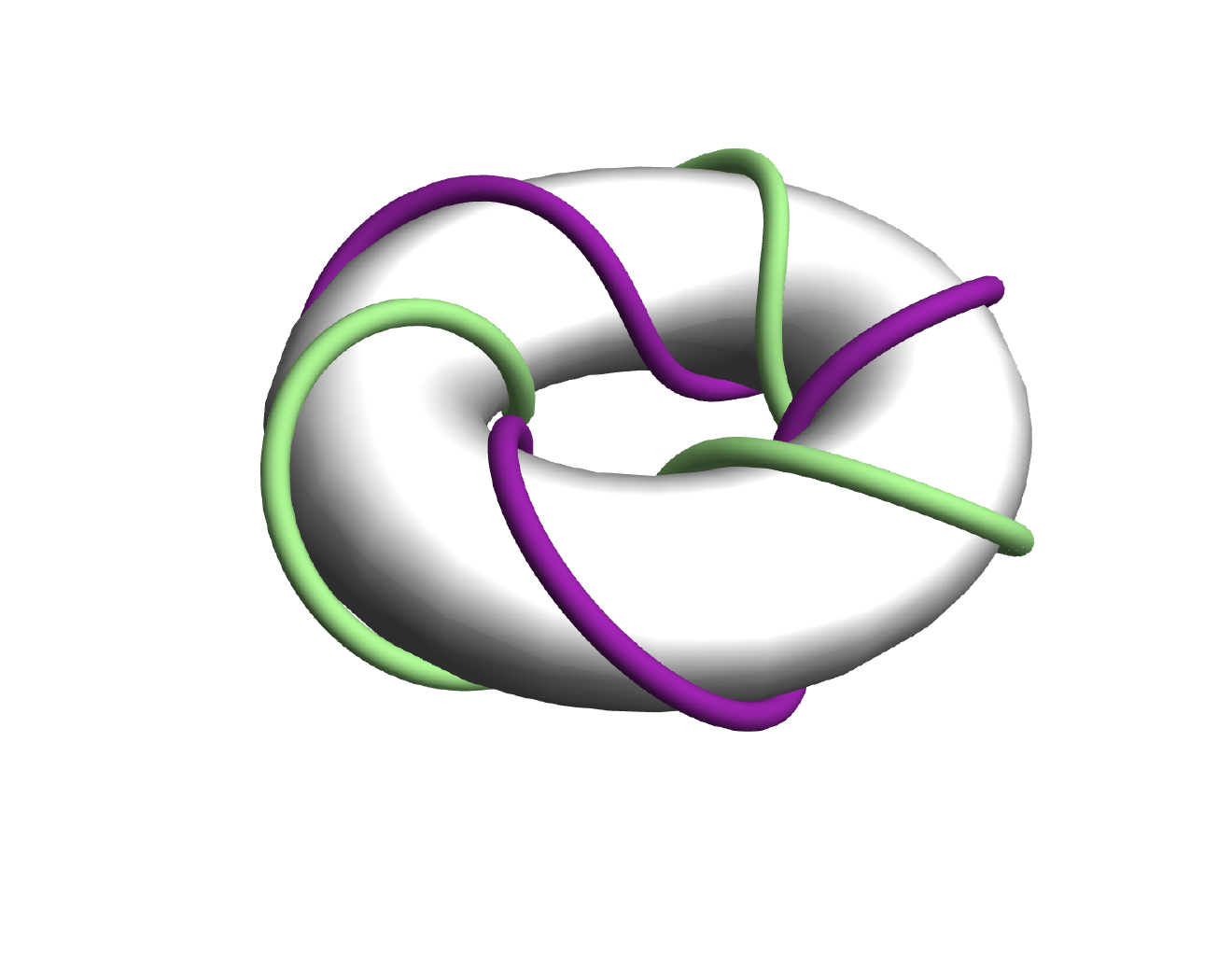}
\caption{A representation of a torus knot (2,3). The wire represents the path followed by the particles on its surface}
\label{Fig:Torus2}
\end{figure}

The sum in Eq. $\left( \ref{eq:Single-Partition-function}\right) $ can be brought to a closed form by using the well-know \textit{Euler-MacLaurin} formula \cite{oliveira2019thermodynamic,oliveira2020thermodynamic}%

\begin{equation}
\sum_{n=0}^{\infty }F\left( n\right) =\int_{0}^{\infty }F\left( n\right)
dn+\frac{1}{2}F\left( 0\right) \displaybreak[0]
 -\frac{1}{2!}B_{2}F^{\prime }\left( 0\right) -\frac{1}{4!}B_{4}F^{\prime
\prime \prime }\left( 0\right) +\ldots ,
\end{equation}%
which allows us to perform the calculation  exactly. The energy spectrum of a particle moving on a torus with dimensions $a,d$ and submitted to a uniform magnetic field that points along the $\boldsymbol{\hat{z}}$ direction is given by \cite{das2016particle}\footnote{It should be noted that we have considered a magnetic field which, in technical terms, is referred as the solenoidal magnetic field (uniform magnetic field pointing in the $\boldsymbol{\hat{z}}$ direction). Nevertheless, for a particle on a torus knot, two other forms of magnetic field, referred
to as toroidal (along the $\boldsymbol{\hat{\phi}}$ direction) and poloidal (along the  $\boldsymbol{\hat{\theta}}$ direction) forms are also relevant. If one is interested in more details ascribed to them, please see Ref. \cite{das2016particle}.}
\begin{equation}
E_{n}=\frac{1}{2m}\left[ \frac{n^{2}\hbar ^{2}}{p^{2}d^{2}}\left( 1-\frac{%
a^{2}q^{2}}{p^{2}d^{2}}\right) +\frac{eB}{p}n\hbar \left( 1+\frac{a^{2}}{%
2d^{2}}-\frac{a^{2}q^{2}}{p^{2}d^{2}}\right) +\frac{e^{2}B^{2}d^{2}}{4}%
\left( 1+\frac{a^{2}}{d^{2}}-\frac{a^{2}q^{2}}{p^{2}d^{2}}\right) \right],
\label{q:SEnergy}
\end{equation}
where $d$ is the outer radius and $a$ the internal radius of the torus respectively with ($d>a$) (see Fig. \ref{Fig:Torus2}). The closed path is characterized by a $(p,q)$-knot with $p$ and $q$ denoting the toroidal and poloidal turns, respectively, so that it turns around $d$ and $a$ respectively. For a nontrivial torus knot, $(p,q)$ are coprime numbers. The well known and simplest knot, a $(2,3)$ or trefoil knot, is depicted in Fig. \ref{Fig:Torus2}.

Now inserting the energy spectrum on the single partition function (\ref{eq:Partition-function}) and computing the sum, we are led to the following single particle partition function:
\begin{eqnarray}
\mathcal{Z}_{1} &=&\sqrt{\frac{\pi mp^{2}d^{2}}{2\beta \left( 1-\mathcal{A}%
\right) }}\exp \left[ \beta \frac{d\alpha^{2}\mathcal{A}}{32m\left( 1-\mathcal{A}%
\right) }\xi \right] \times  \notag \\
&\phantom{=}&\times \left\{ 1+ \mathrm{erf} \left[ -\frac{\xi }{4}\sqrt{\frac{2\beta }{%
m\left( 1-\mathcal{A}\right) }}\left( 1-\mathcal{A}+\frac{\alpha^{2}}{2}\mathcal{A}%
\right) \right] \right\} ,  \label{eq:Exact_Z}
\end{eqnarray}%
where $\mathrm{erf}$ is the error function given by
\begin{equation}
\mathrm{erf}\left[ z\right]  =\frac{2}{\sqrt{\pi }}\int_{0}^{z}e^{-t^{2}}dt,
\end{equation}%
and we also have defined the following quantities%
\begin{equation}
\xi =eB,\phantom{ss}\mathcal{A}=\frac{a^{2}}{d^{2}},\phantom{ss}\alpha=\frac{q}{p},
\end{equation}%
where $\mathcal{A}$ is a property of the torus whereas $\alpha$ characterizes the knot of the particular path. The quantity $\alpha$ is called the “winding number” of the $(p, q)$-torus
knot, and it is also a simple way of measuring the complexity of the knot. Notice that two different torus knots can have the same $\alpha$. Nevertheless, $\alpha$ will be  different if  at least one of $p$ or $q$ differ. In this sense, $\alpha$ is regarded as a “unique”
identity of a $(p, q)$-torus knot. It is worth mentioning that similar forms of partition function have appeared in very recent works in the literature considering different contexts \cite{araujo2021thermodynamic,araujo2021lorentz,petrov2021higher,petrov2021bouncing,aa2021thermodynamics22}.

Now substituting $\left( \ref{eq:Exact_Z}\right) $ in $\left( \ref{eq:Helmm-f}%
\right) $, we find%
\begin{eqnarray}
f &=&-\frac{d\alpha^{2}\mathcal{A}}{32m\left( 1-\mathcal{A}\right) }\xi -\frac{1}{%
\beta }\ln \sqrt{\frac{\pi mp^{2}d^{2}}{2\beta \left( 1-\mathcal{A}\right) }}
\notag \\
&\phantom{=}&-\frac{1}{\beta }\ln \left\{ 1+\mathrm{erf}\left[ -\frac{\xi }{4}\sqrt{\frac{%
2\beta }{m\left( 1-\mathcal{A}\right) }}\left( 1-\mathcal{A}+\frac{\alpha^{2}}{2}\mathcal{A}\right) \right] \right\} .
\end{eqnarray}%
From this expression, using previous relations (\ref{eq:Therm-functions1}), we can get the relevant thermodynamic variables. For
instance, the internal energy is given by%
\begin{eqnarray}
u &=&-\frac{1}{\beta ^{2}}\left( \frac{1}{2}+\ln \sqrt{\frac{\pi mp^{2}d^{2}%
}{2\beta \left( 1-\mathcal{A}\right) }}\right)  \notag \\
&&-\frac{1}{\beta ^{2}}\ln \left\{ 1+\mathrm{erf}\left[ -\frac{\xi }{4}\sqrt{%
\frac{2\beta }{m\left( 1-\mathcal{A}\right) }}\left( 1-\mathcal{A}+\frac{%
\alpha^{2}}{2}\mathcal{A}\right) \right] \right\}  \notag \\
&&-\frac{\xi }{\beta ^{\frac{3}{2}}}\sqrt{\frac{1}{8m\pi \left( 1-\mathcal{A}%
\right) }}\frac{\left( 1-\mathcal{A}+\frac{\alpha^{2}}{2}\mathcal{A}\right) \exp \left[
-\frac{\xi ^{2}}{8}\frac{\beta }{m\left( 1-\mathcal{A}\right) }\left( 1-%
\mathcal{A}+\frac{\alpha^{2}}{2}\mathcal{A}\right) ^{2}\right] }{1-\mathrm{erf}\left[
\frac{\xi }{4}\sqrt{\frac{2\beta }{m\left( 1-\mathcal{A}\right) }}\left( 1-\mathcal{A}+\frac{\alpha^{2}}{2}\mathcal{A}\right)\right]}.
\end{eqnarray}%
Another important quantity is the magnetization, which reads
\begin{equation}
\mathfrak{m}=\frac{d\alpha^{2}\mathcal{A}}{32m\left( 1-\mathcal{A}\right) }-\sqrt{\frac{1}{2\pi
\beta m\left( 1-\mathcal{A}\right) }}\frac{\left( 1-\mathcal{A}+\frac{%
\alpha^{2}}{2}\mathcal{A}\right) \exp \left[ -\frac{\xi ^{2}}{8}\frac{\beta }{m\left(
1-\mathcal{A}\right) }\left( 1-\mathcal{A}+\frac{\alpha^{2}}{2}\mathcal{A}\right) ^{2}%
\right] }{1-\mathrm{erf}\left[ \frac{\xi }{4}\sqrt{\frac{2\beta }{m\left( 1-%
\mathcal{A}\right) }}\left( 1-\mathcal{A}+\frac{\alpha^{2}}{2}\mathcal{A}\right) %
\right] }.
\end{equation}%
The susceptibility can also be calculated and is given by%
\begin{equation}
\chi =-\frac{4}{\pi \beta }\frac{\mathcal{B}^{2}\exp \left( -2\mathcal{B}%
^{2}\xi ^{2}\right) }{1-\mathrm{erf}\left( \mathcal{B}\xi \right) }\left\{
\frac{1}{1-\mathrm{erf}\left( \mathcal{B}\xi \right) }-\sqrt{\pi }\mathcal{B}%
\xi \exp \left( \mathcal{B}^{2}\xi ^{2}\right) \right\} ,
\end{equation}%
where we have used the shorthand notation%
\begin{equation}
\mathcal{B}=\frac{1}{4}\sqrt{\frac{2\beta }{m\left( 1-\mathcal{A}\right) }}%
\left( 1-\mathcal{A}+\frac{\alpha ^{2}}{2}\mathcal{A}\right) .
\end{equation}%
Since expressions of  energy,  magnetization and  susceptibility are quite lengthy, instead of providing their explicit forms, we show their behavior pictorially. Similarly, in this way, for  entropy and  heat capacity in the low energy regime, we shall follow the same procedure. Let us explore some interesting features from the magnetization and the susceptibility for a particular configuration of magnetic field and temperature. Initially, we consider the configuration in which $\xi =0$. With this, we obtain the following results:
\begin{align}
\mathfrak{m}_{\xi =0}& =\frac{d\alpha ^{2}\mathcal{A}}{32m\left( 1-\mathcal{A%
}\right) }-\frac{1-\mathcal{A}+\frac{\alpha ^{2}}{2}\mathcal{A}}{\sqrt{2\pi
\beta m\left( 1-\mathcal{A}\right) }}, \\
\chi _{\xi =0}& =-\frac{4\mathcal{B}^{2}}{\pi \beta }.
\end{align}%
Now, for both $\xi =0$ and $T=0$, we get%
\begin{align}
\mathfrak{m}_{\xi =0,T=0}& =\frac{d\alpha ^{2}\mathcal{A}}{32m\left( 1-%
\mathcal{A}\right) }, \\
\chi _{\xi =0,T=0}& =-\frac{1}{2\pi }\frac{1}{m\left( 1-\mathcal{A}\right) }%
\left( 1-\mathcal{A}+\frac{\alpha ^{2}}{2}\mathcal{A}\right) ^{2}.
\end{align}%
It is interesting to note that in the above restricted cases, the topology of the path appears only through $\alpha$ whereas both $a,d$ appear to represent geometry of the path. Furthermore, the residual magnetization and susceptibility are mainly characterized by its geometrical and topological aspects. For instance, if the topological parameter $\alpha $ increases, then both $\mathfrak{m}_{\xi =0,T=0}$ and $\chi _{\xi =0,T=0}$
increase as well their magnitude. Therefore, the residual magnetic properties will depend on the complexities of the path that the electrons follow.

\begin{figure}[h!]
\centering
\includegraphics[width=8cm,height=5cm]{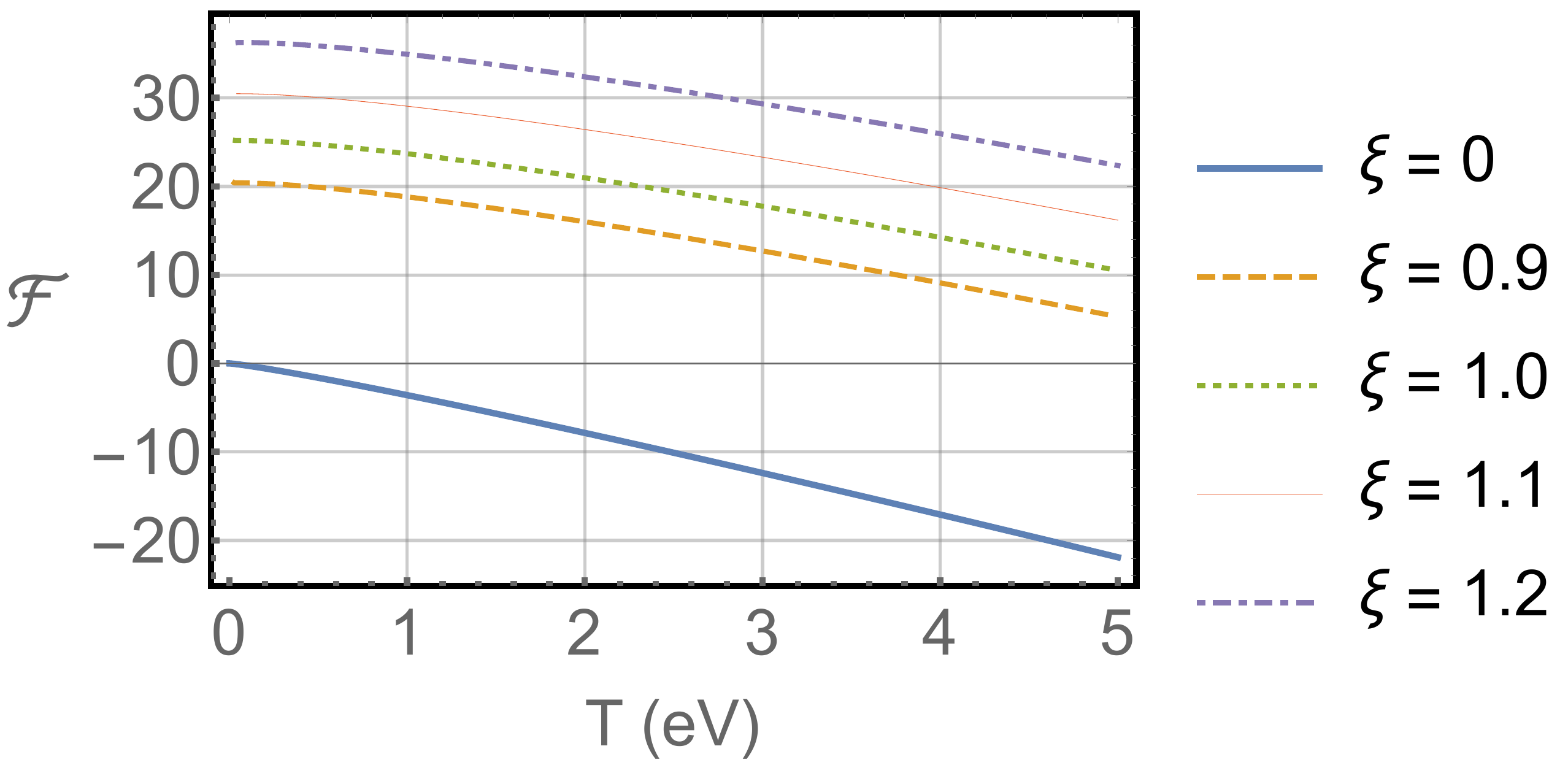}
\includegraphics[width=8cm,height=5cm]{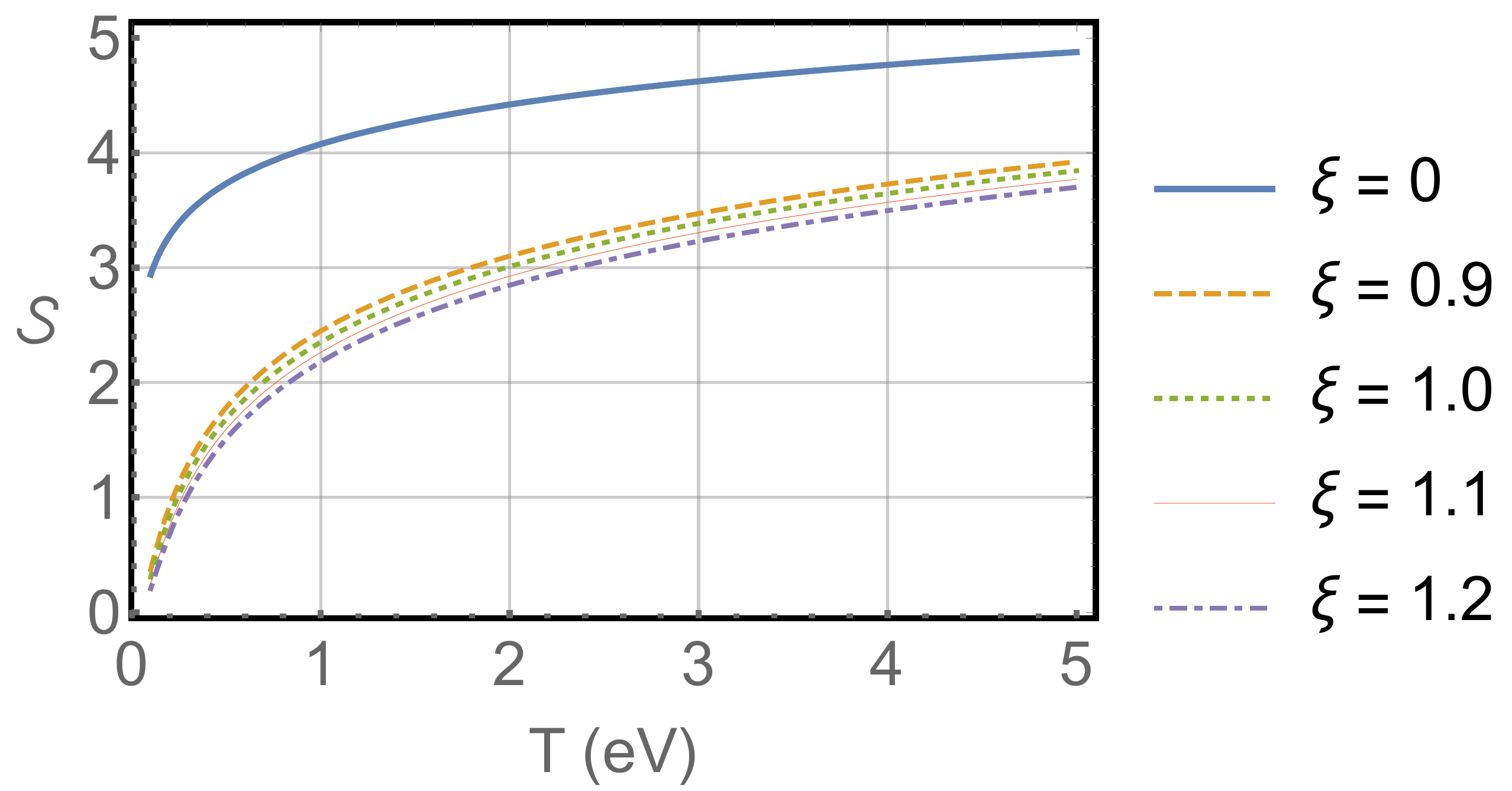}
\caption{Helmholtz free energy $({\cal{F}})$ and entropy $({\cal{S}})$ versus temperature ($T$)}
\label{Fig:Fenergy}
\end{figure}

\begin{figure}[h!]
\centering
\includegraphics[width=8cm,height=5cm]{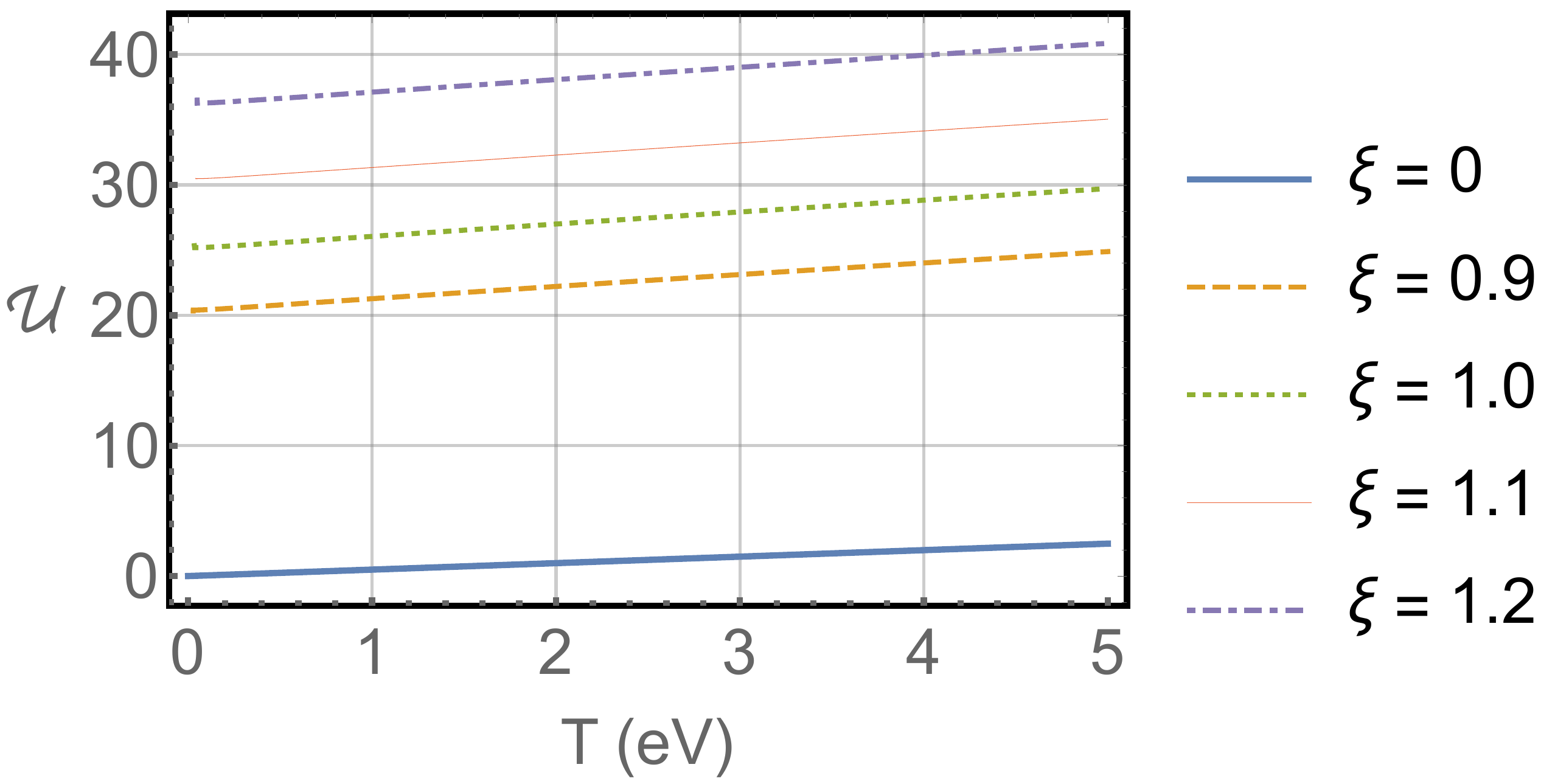}
\includegraphics[width=8cm,height=5cm]{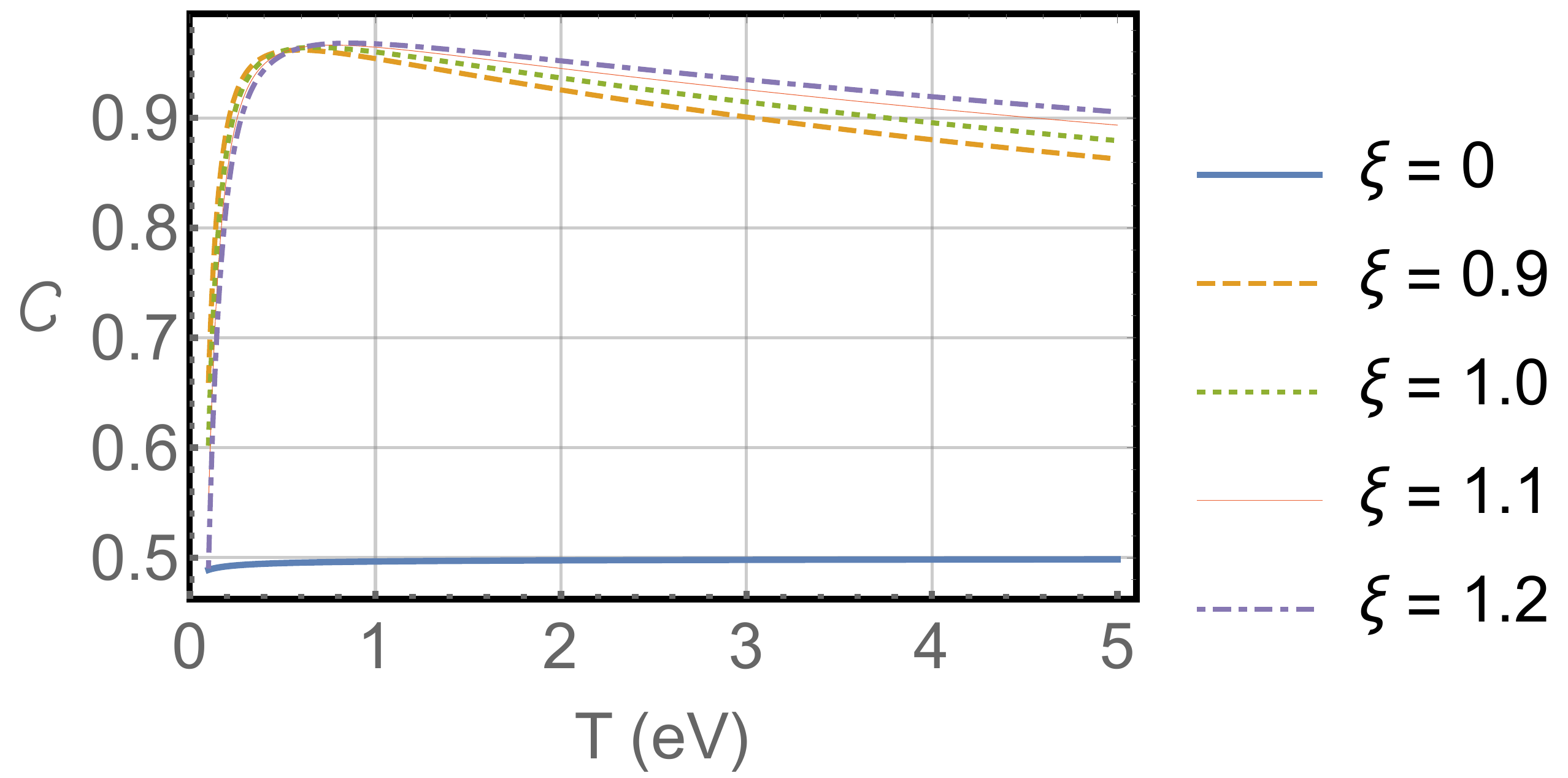}
\caption{Internal energy $({\cal{U}})$ and Heat capacity $({\cal{C}})$ versus temperature ($T$)}
\label{Fig:Uenergy}
\end{figure}

\begin{figure}[h!]
\centering
\includegraphics[width=8cm,height=5cm]{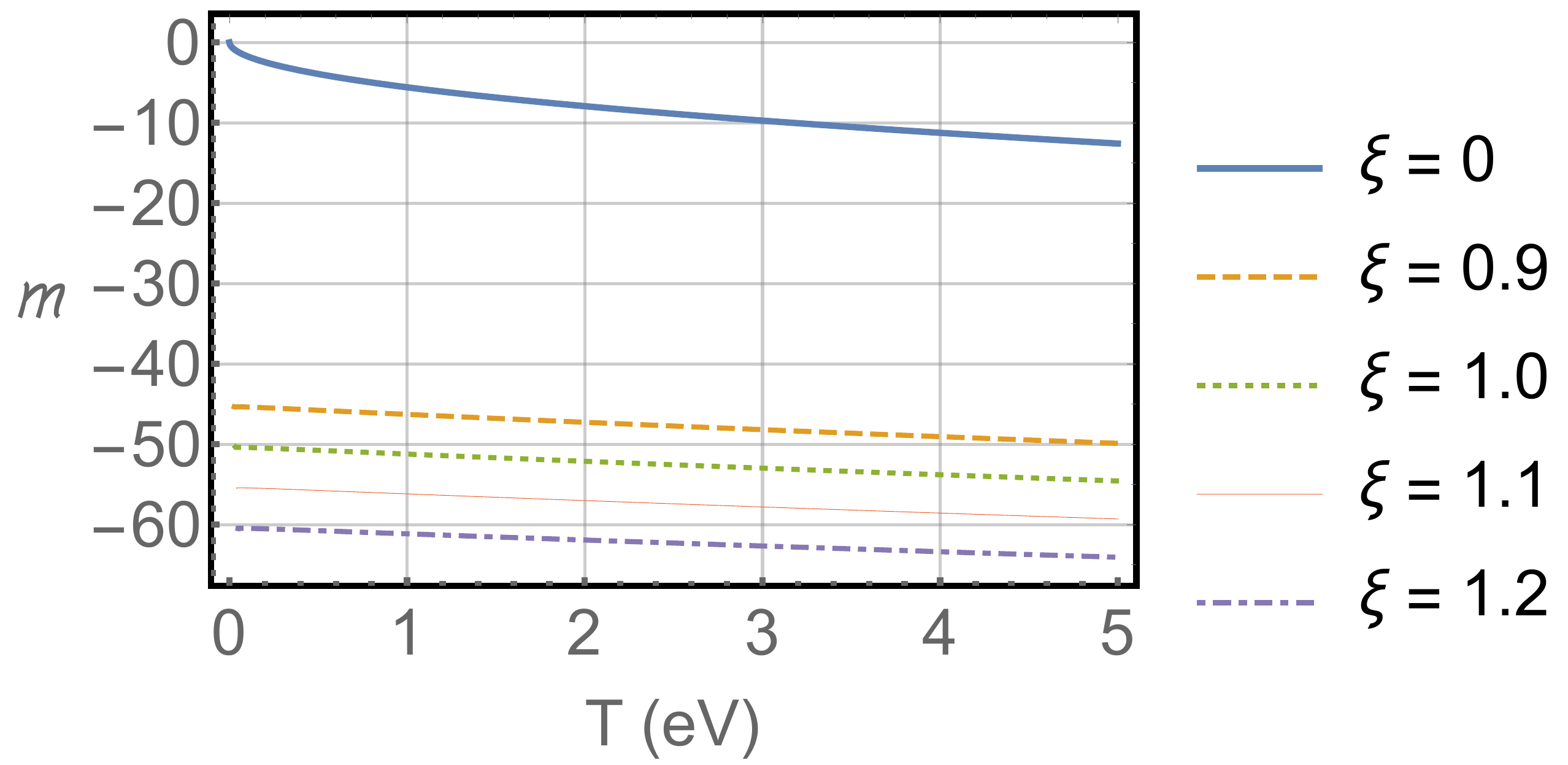}
\includegraphics[width=8cm,height=5cm]{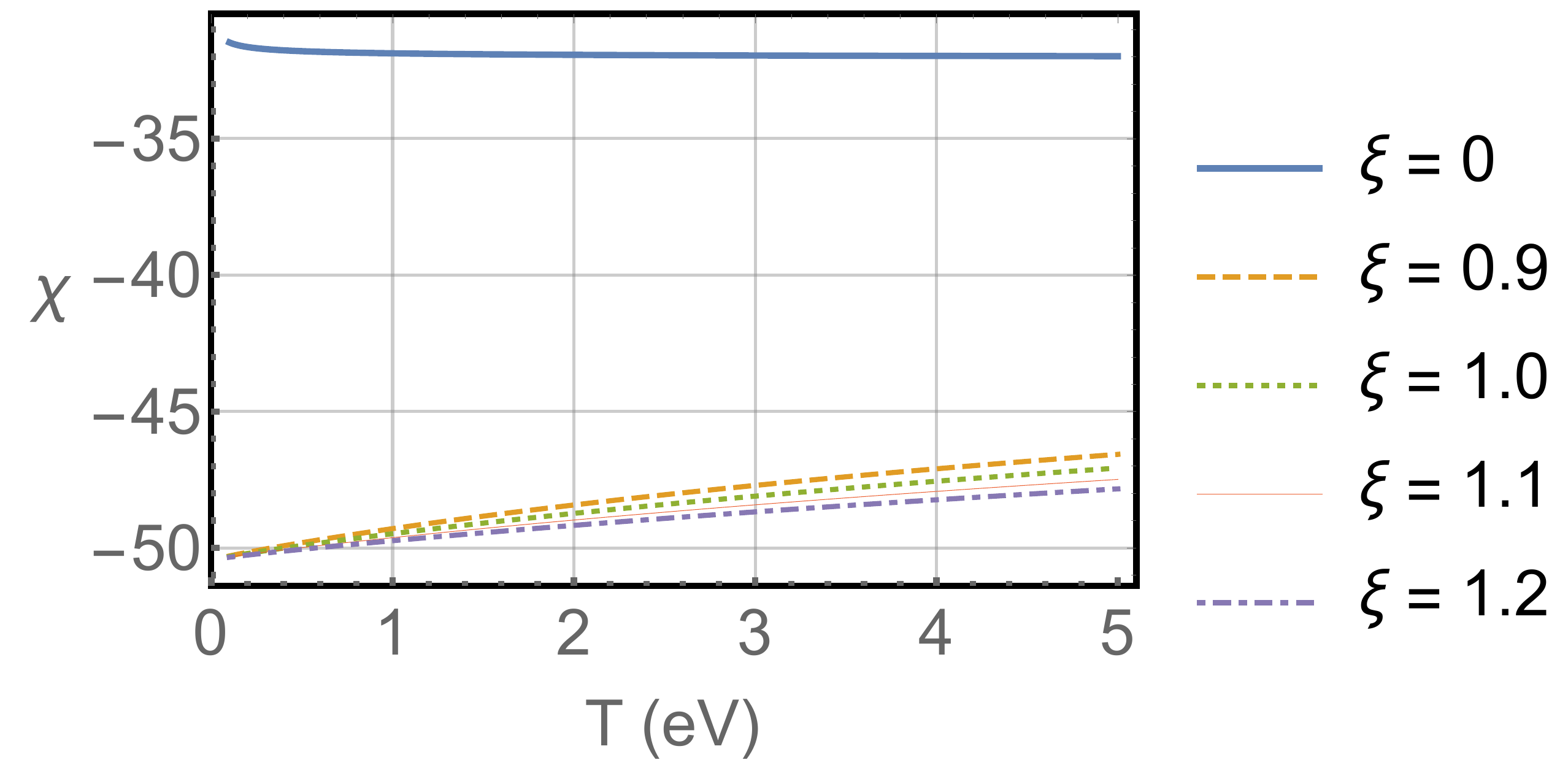}
\caption{Magnetization $({\cal{M}})$ and Susceptibility $({\cal{\chi}})$ versus temperature ($T$)}
\label{Fig:Magnetization}
\end{figure}

In Figs. \ref{Fig:Fenergy}, \ref{Fig:Uenergy} and \ref{Fig:Magnetization}, we have used the values $p=2$ and $q=3$, so that the first coprimes yielding  the simplest and widely studied $(2,3)$-knot or the trefoil knot.  The plots below exhibit the behavior of the thermodynamic functions with respect to the temperature. Moreover,  we also compare the behavior of those functions for different values of the parameter $\xi$, which controls the intensity of the external magnetic field. Since our system consists of  electrons,  we need to take its mass into account in our numerical calculations.

\begin{figure}[h!]
\centering
\includegraphics[scale=0.35]{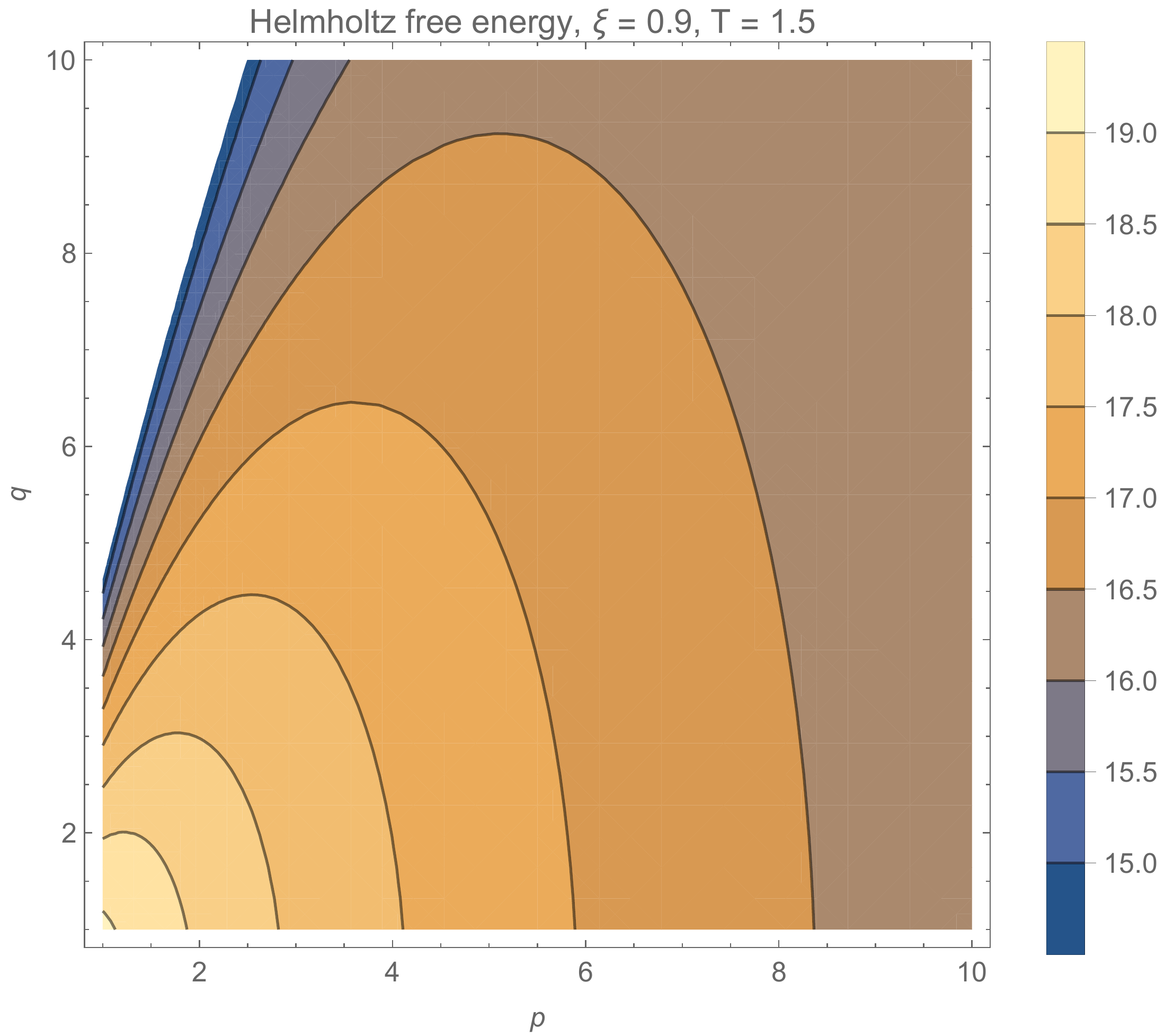}
\includegraphics[scale=0.35]{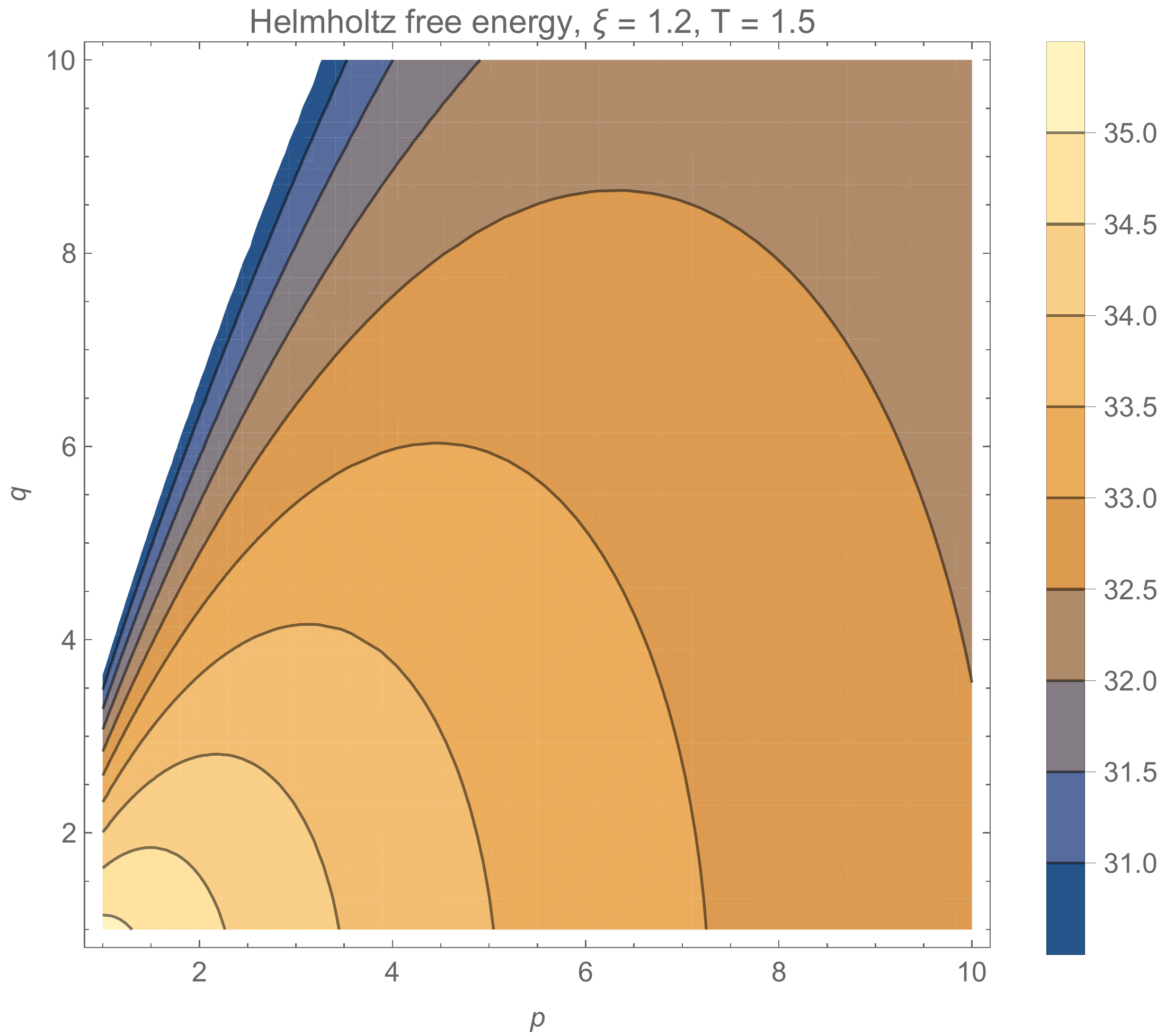}\\
\includegraphics[scale=0.35]{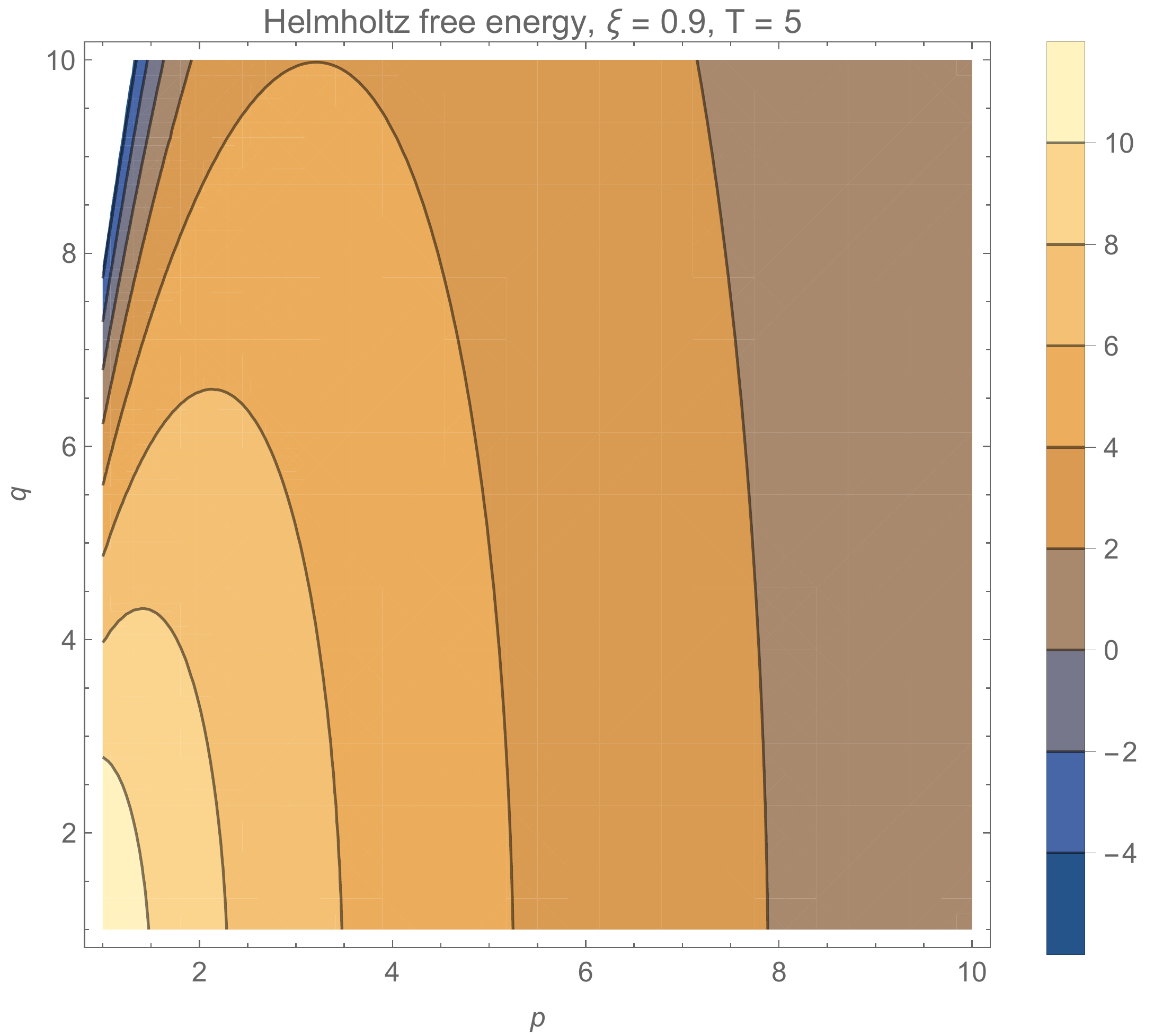}
\includegraphics[scale=0.35]{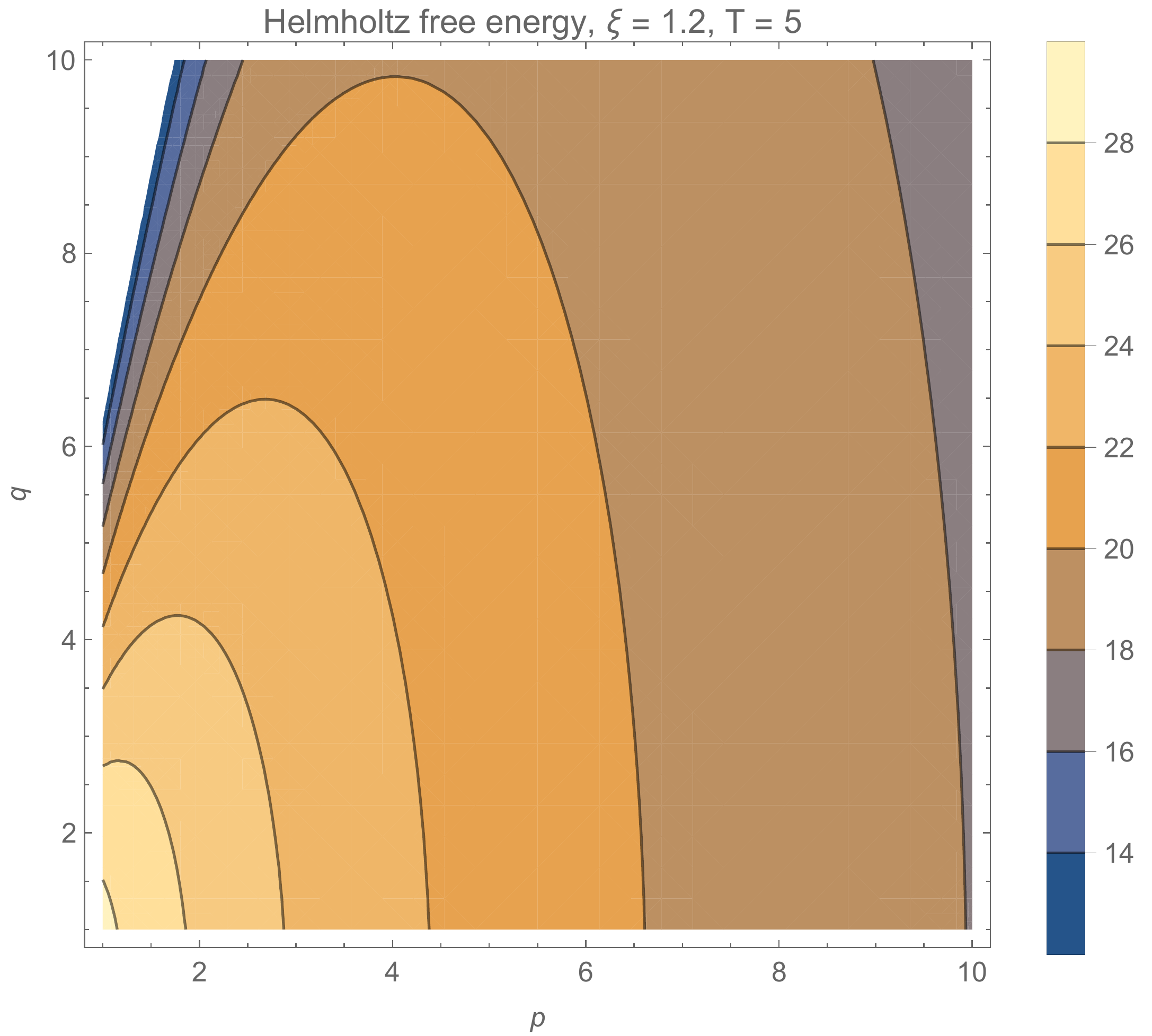}
\caption{These plots exhibit the Helmholtz free energy for different configurations of temperature and magnetic field. It  also displays the behavior of  Helmholtz energy as a function of the wind numbers $(p,q)$. Note that  Free energy increases when  magnetic field increases and also  when both  $p$ and $q$ increases.}
\label{fig:Helmholtz3D}
\end{figure}

\begin{figure}[h!]
\centering
\includegraphics[scale=0.35]{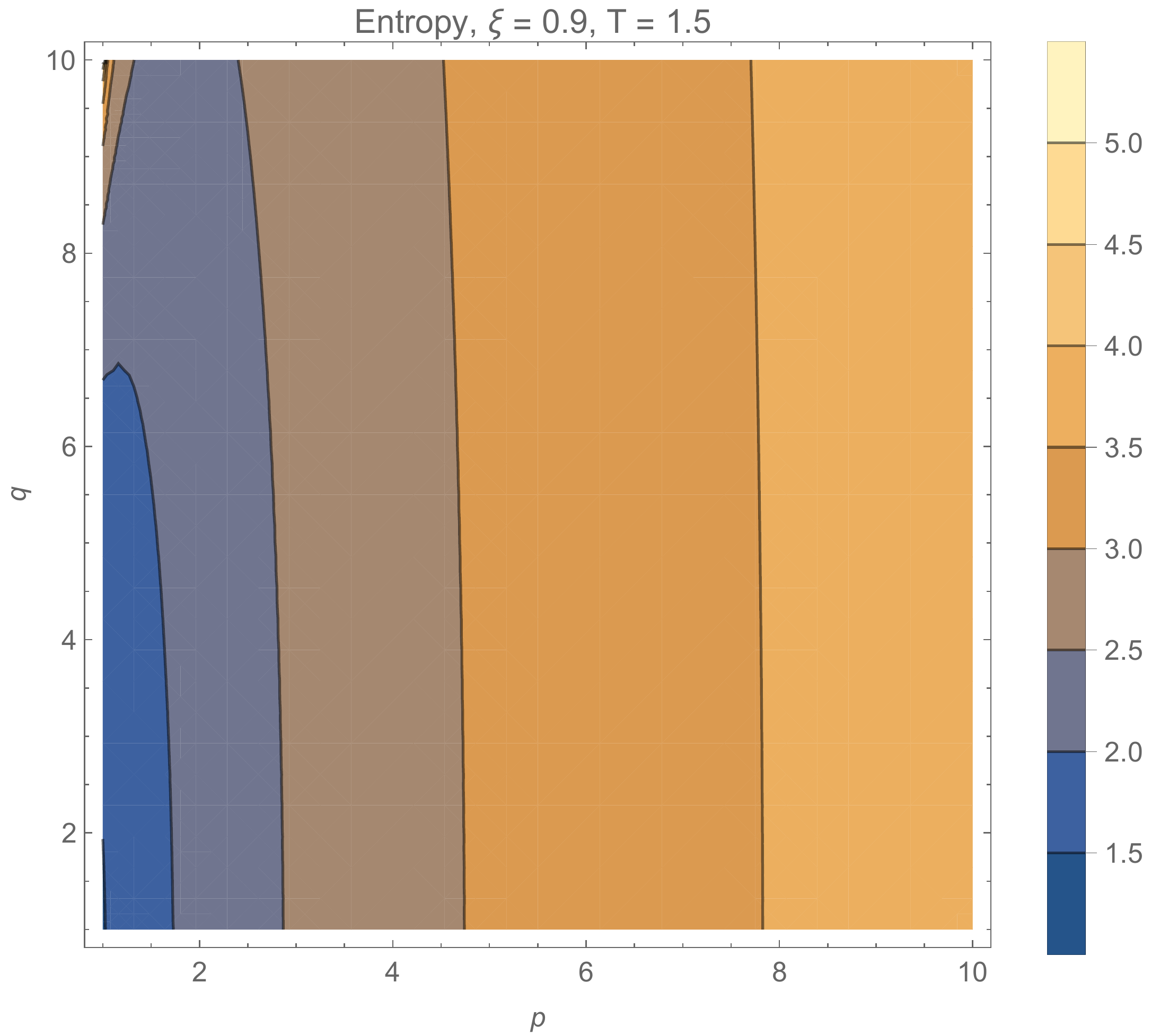}
\includegraphics[scale=0.35]{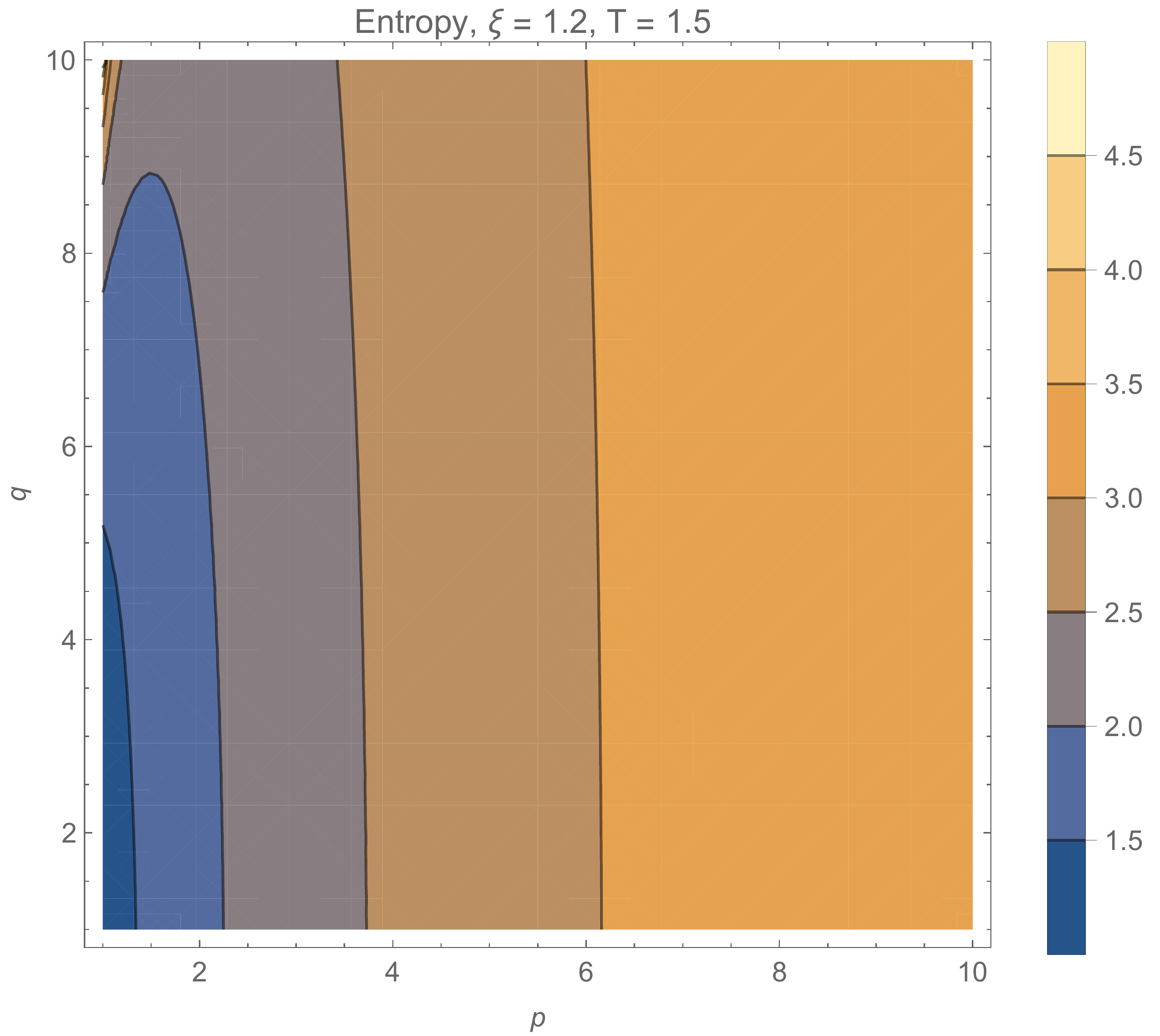}\\
\includegraphics[scale=0.35]{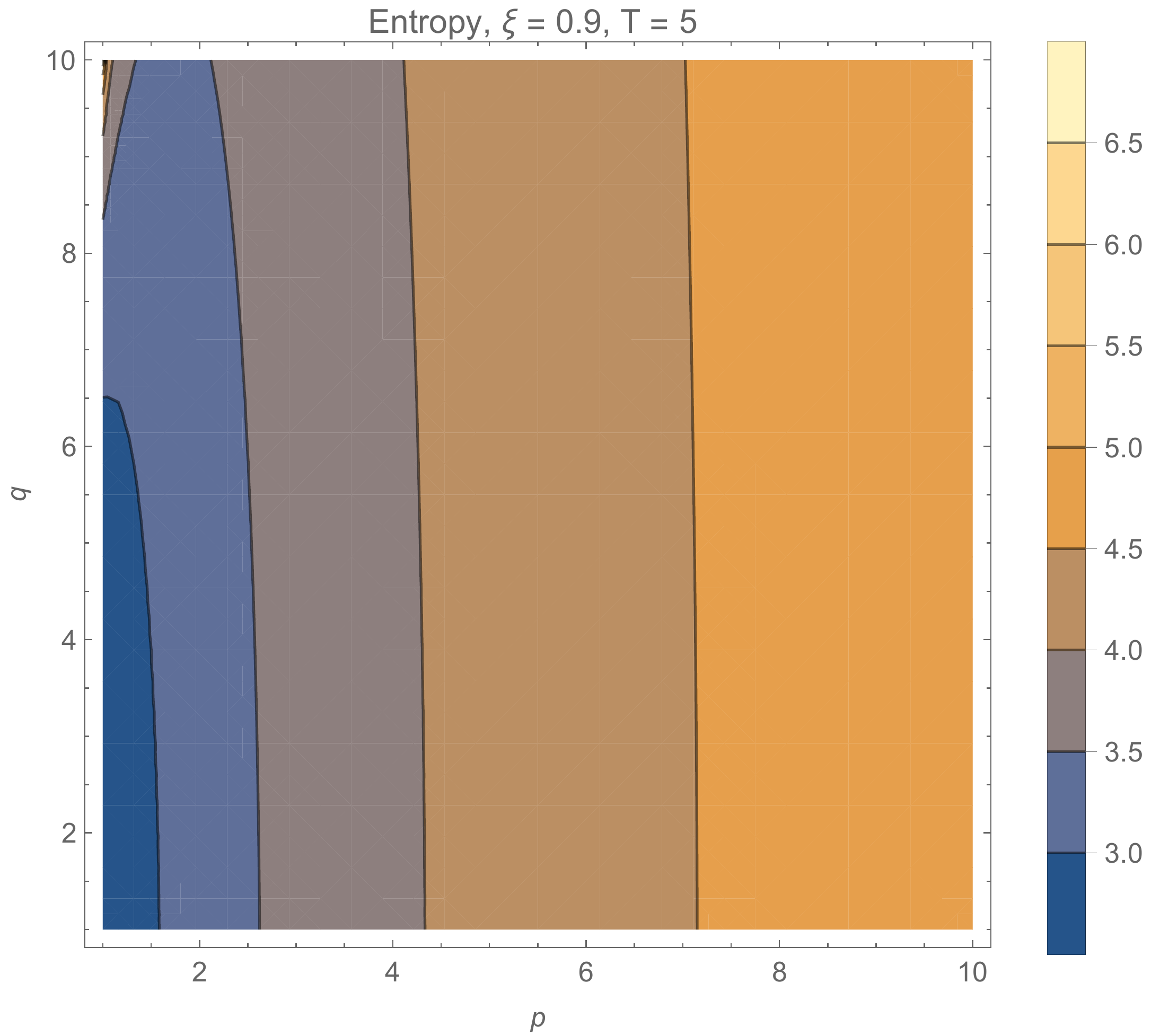}
\includegraphics[scale=0.35]{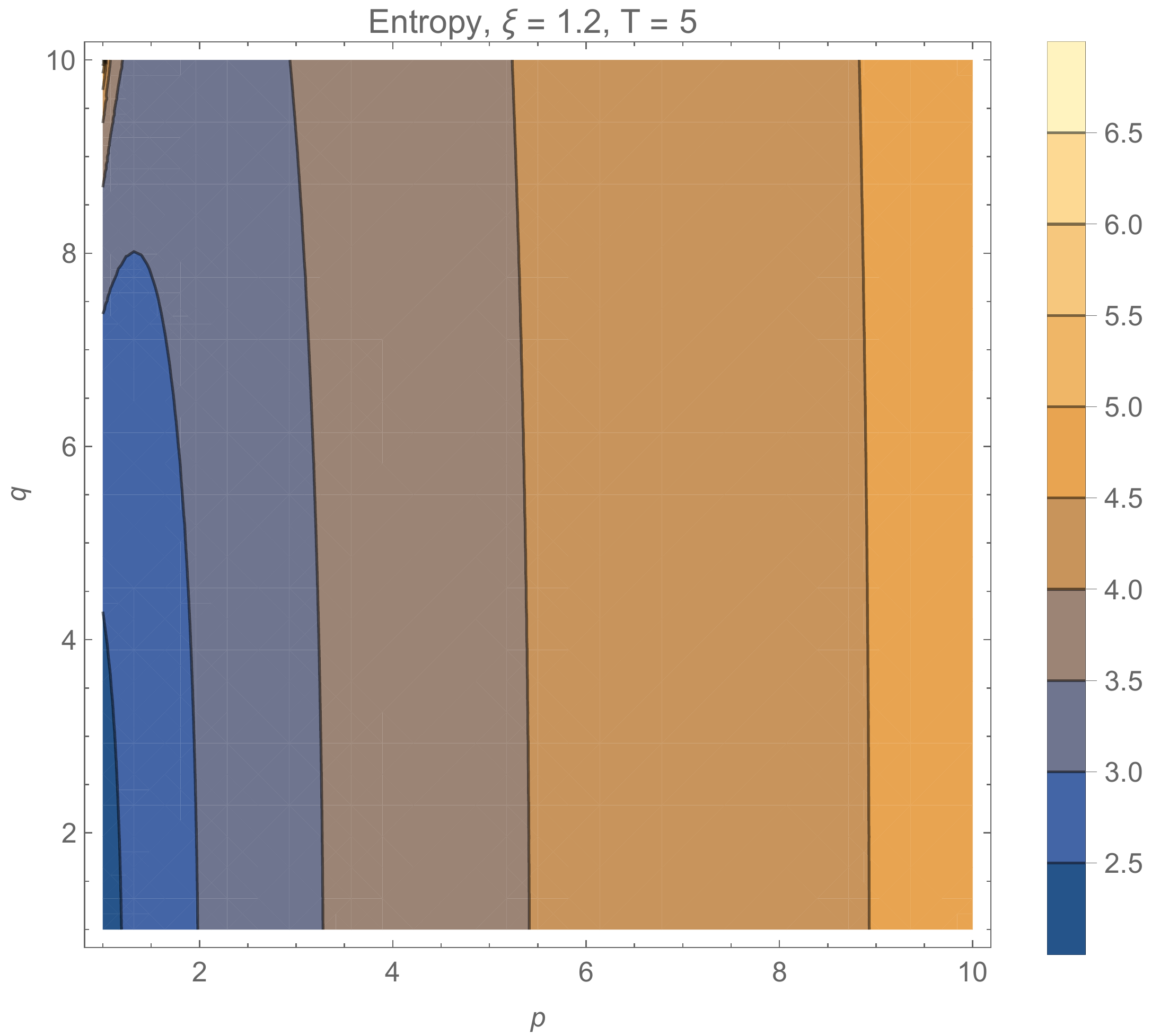}
\caption{These plots exhibit the entropy for different configurations of temperature and magnetic field. We also see its behavior as a function of the wind numbers $(p,q)$. The behavior of the entropy is similar to the free energy, i.e. it increases with the parameters $p$ and $q$.}
\label{fig:Entropy3D}
\end{figure}

\begin{figure}[h!]
\centering
\includegraphics[scale=0.35]{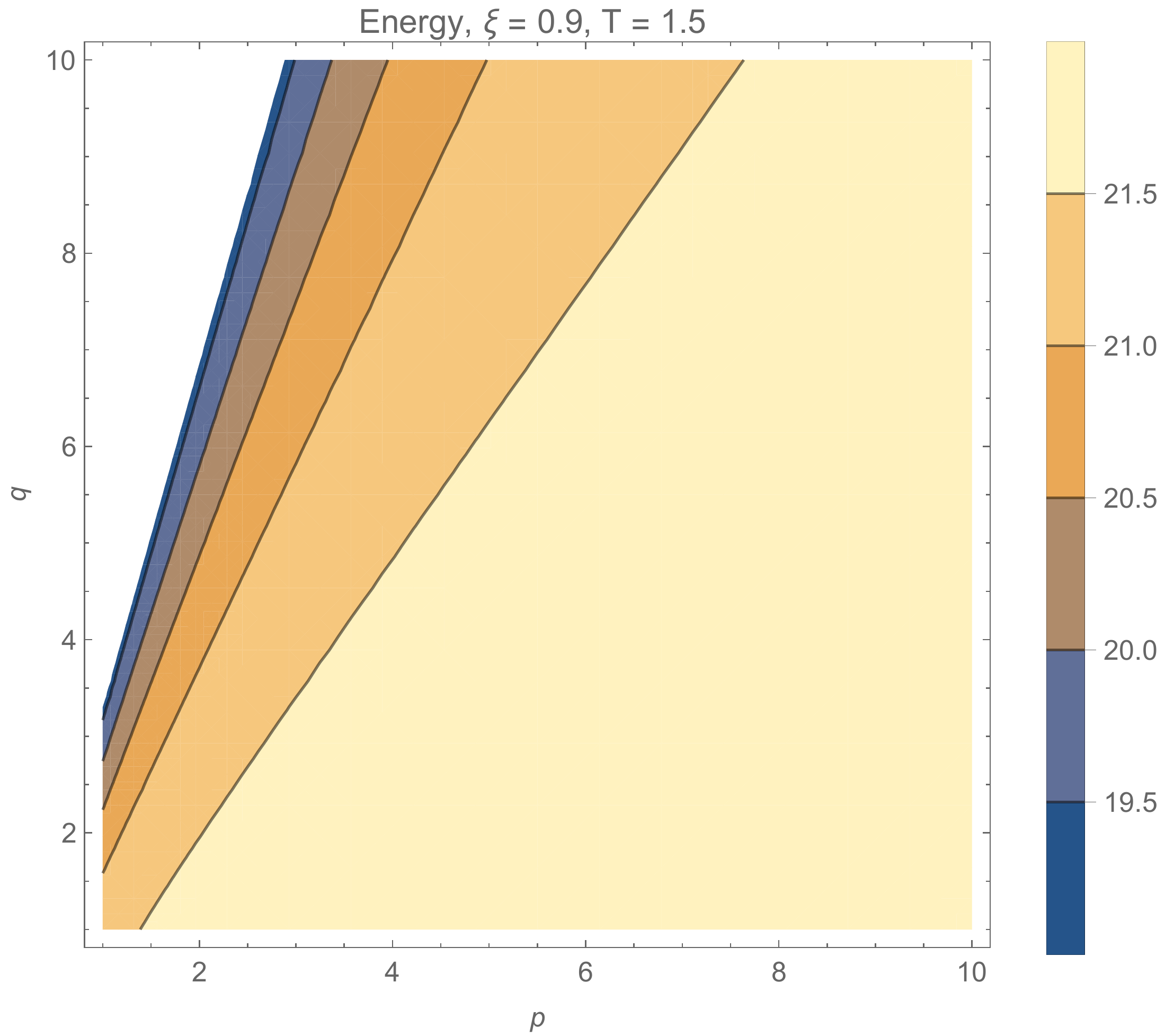}
\includegraphics[scale=0.35]{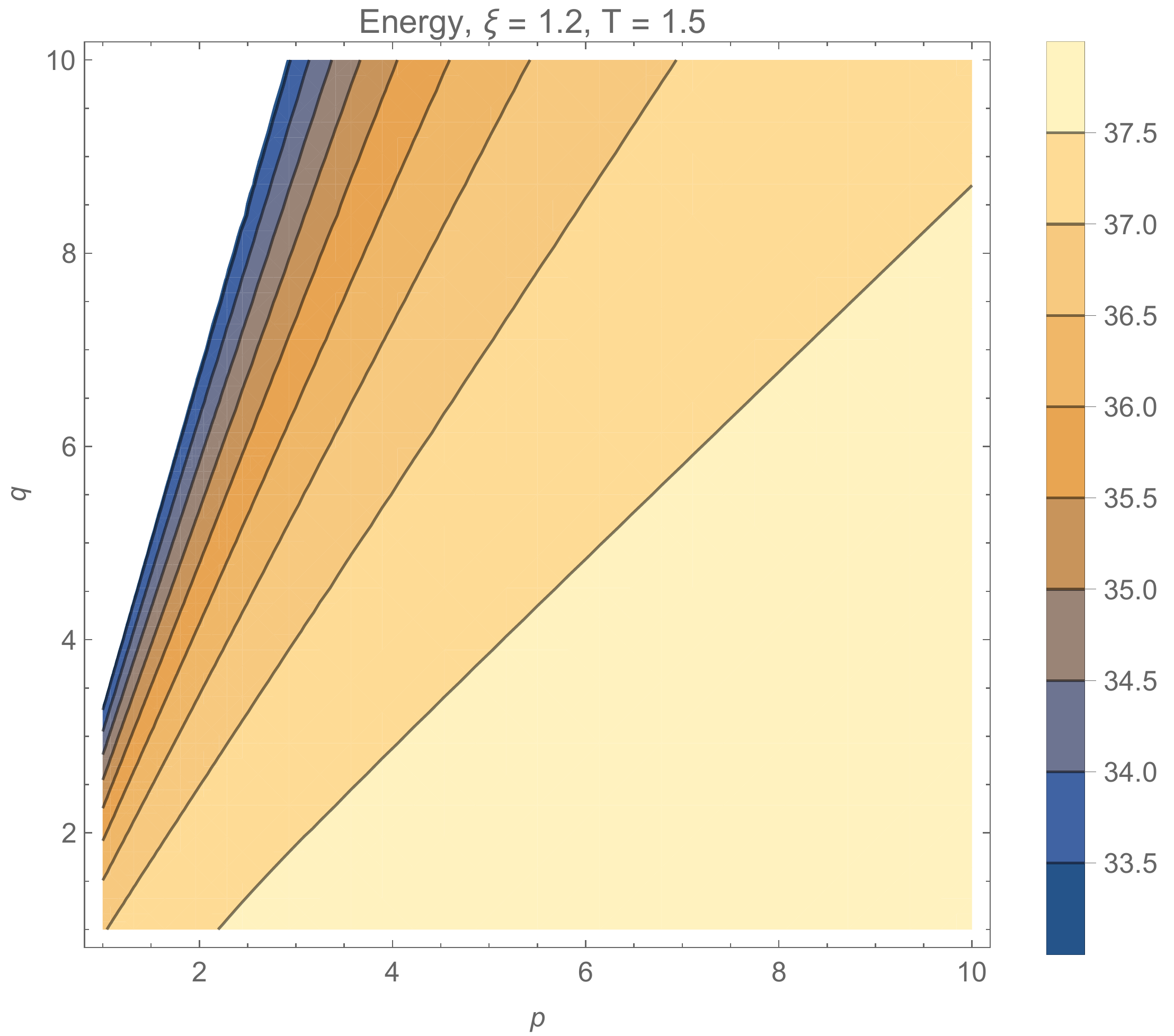}\\
\includegraphics[scale=0.35]{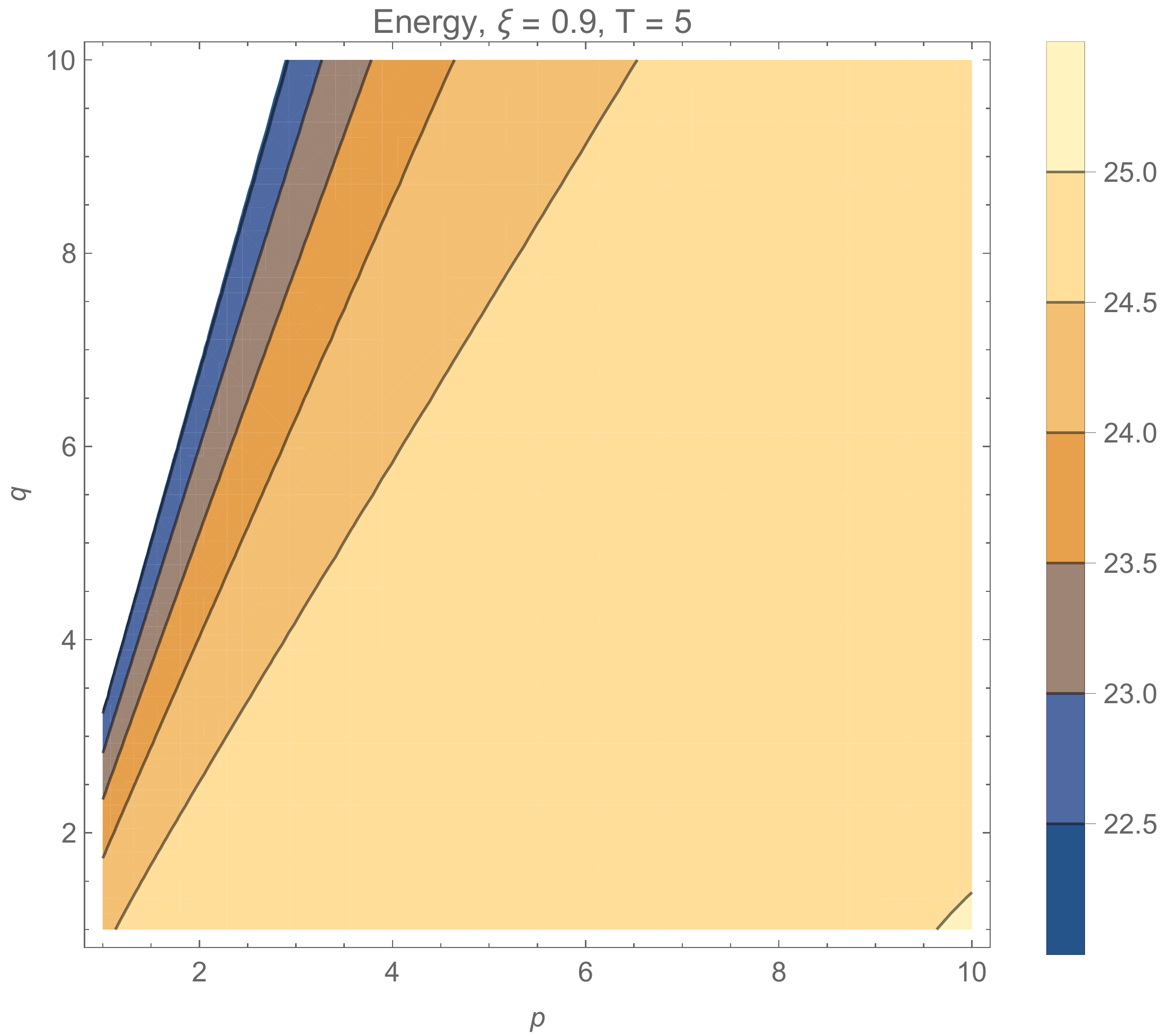}
\includegraphics[scale=0.35]{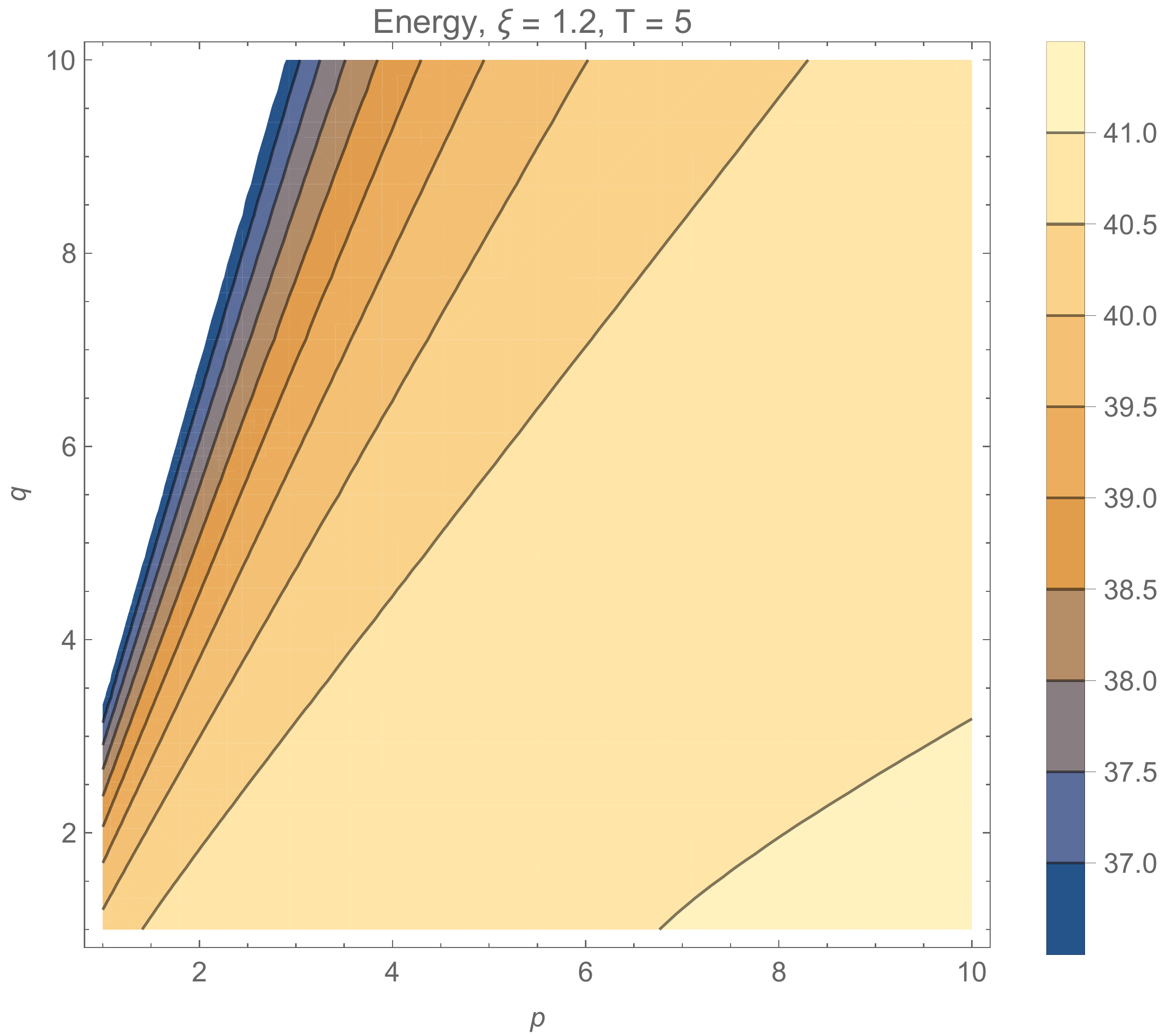}
\caption{These plots exhibit the contour plot for the mean energy for different configurations of temperature and magnetic field. In those plots we can see the behavior of the Energy as a function of the wind numbers $(p,q)$.}
\label{fig:Energy3D}
\end{figure}

\begin{figure}[h!]
\centering
\includegraphics[scale=0.35]{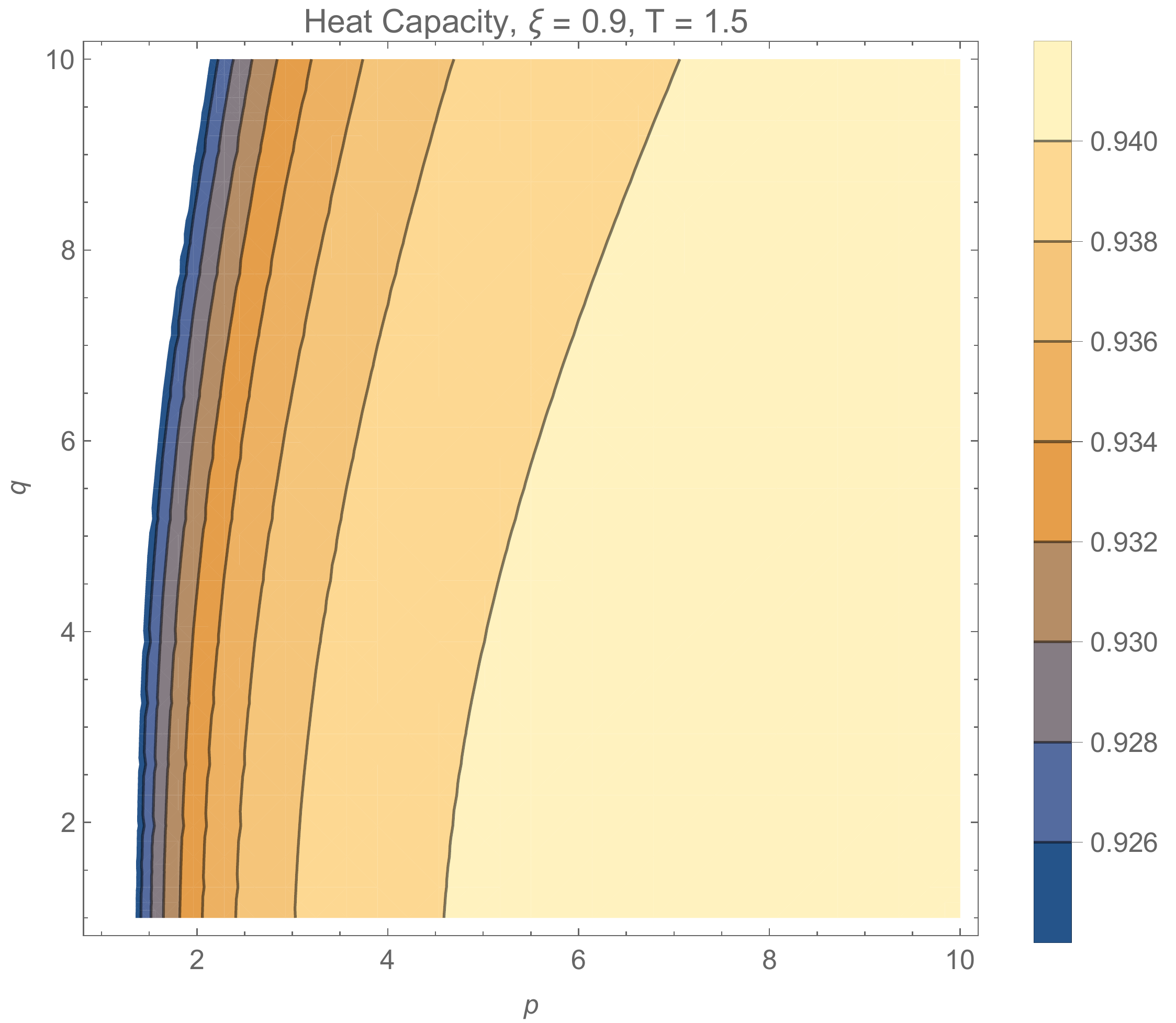}
\includegraphics[scale=0.35]{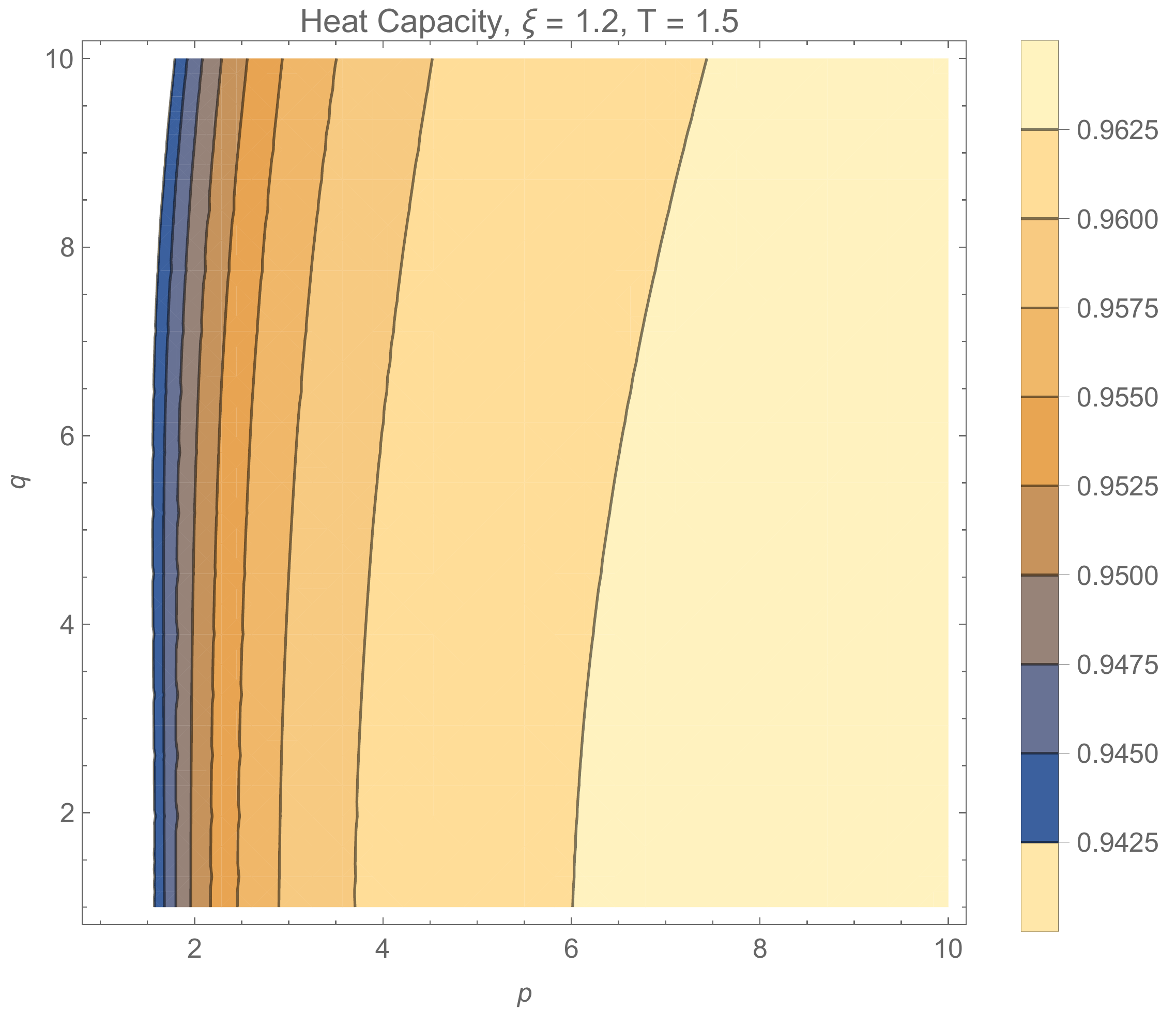}\\
\includegraphics[scale=0.35]{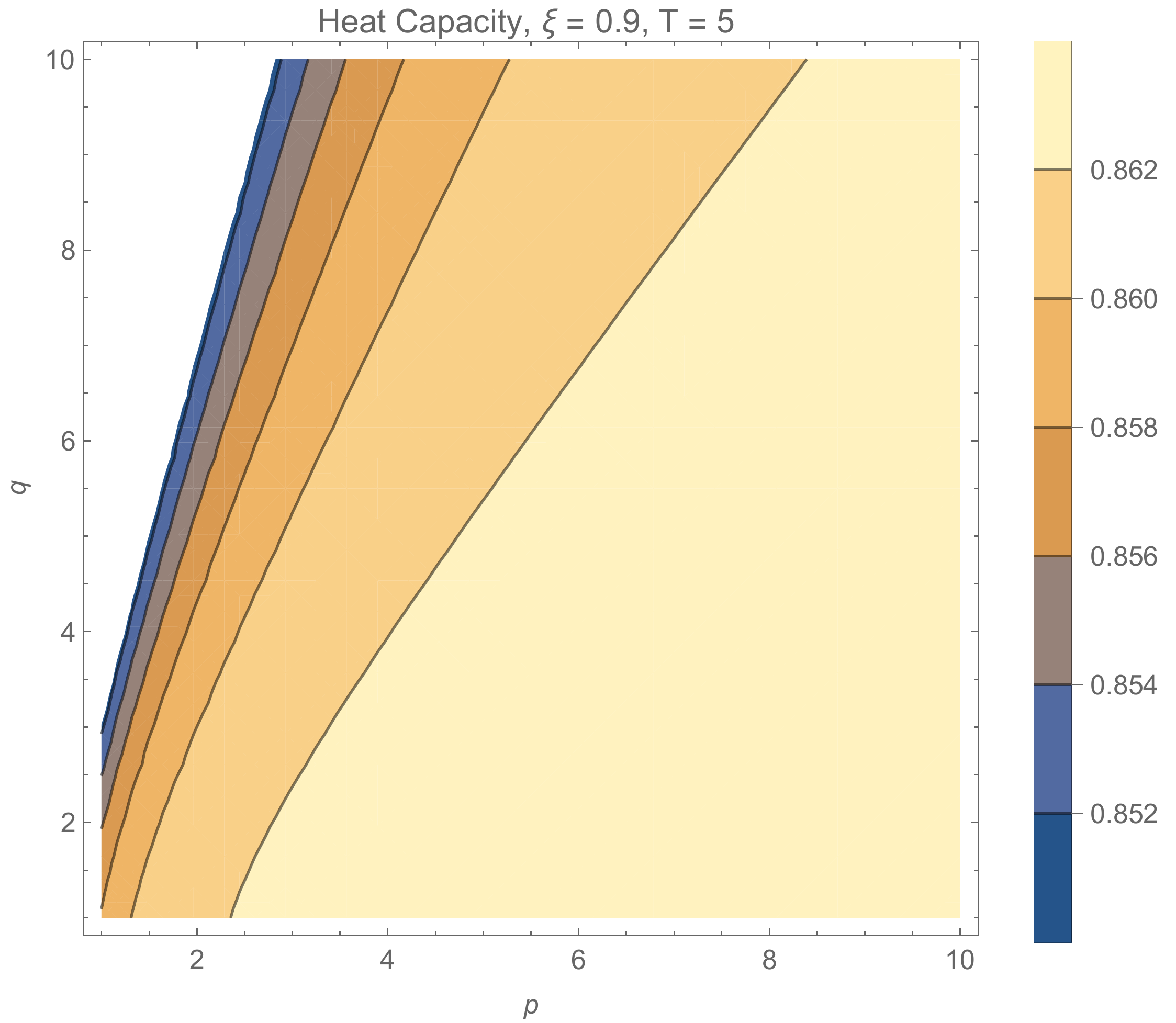}
\includegraphics[scale=0.35]{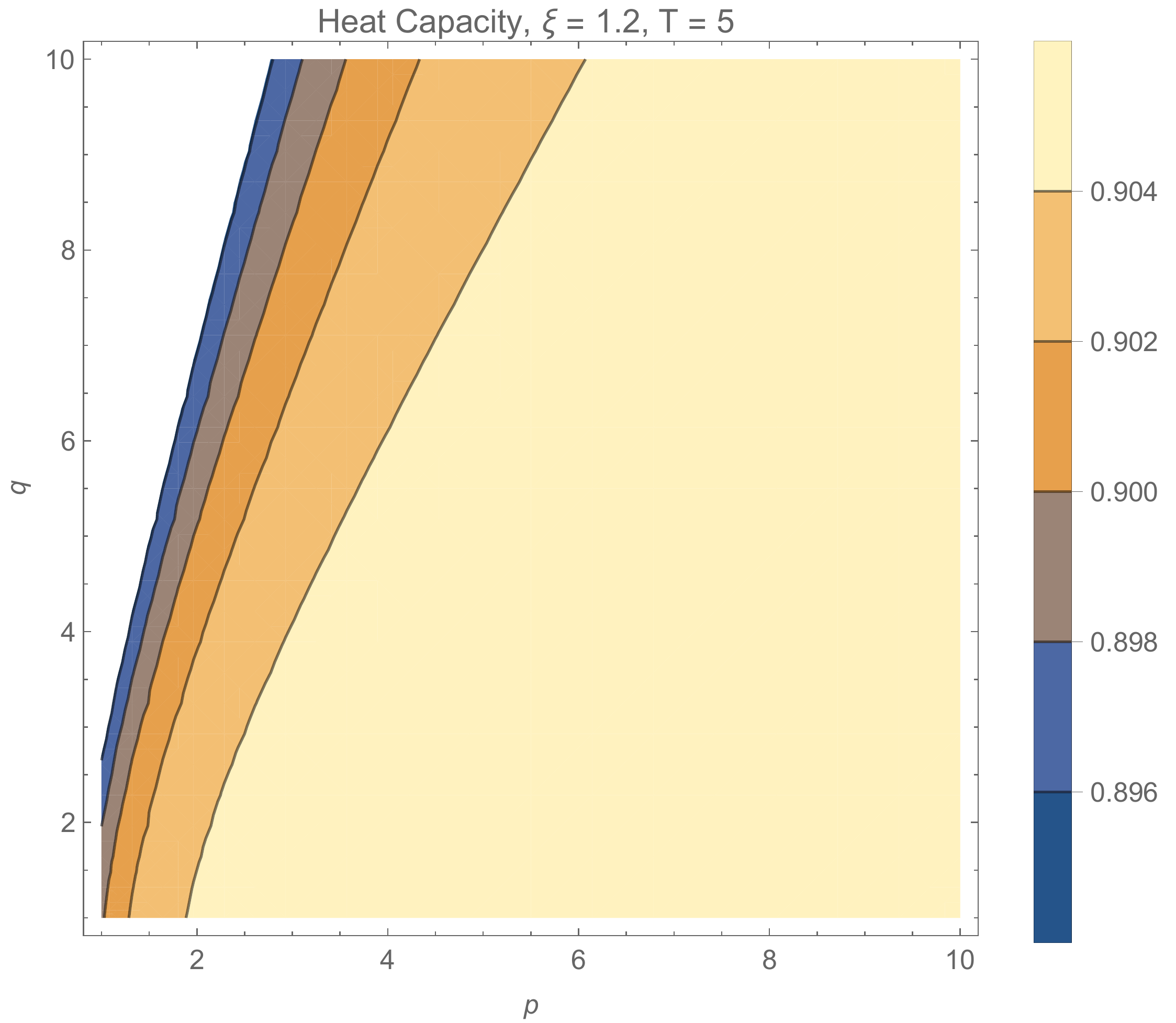}
\caption{The contour plots for the Heat Capacity for different configurations of temperature and magnetic field. In these ones, we can see the behavior of the Heat Capacity as a function of the wind numbers $(p,q)$.}
\label{fig:HCapacity3D}
\end{figure}

\begin{figure}[h!]
\centering
\includegraphics[scale=0.35]{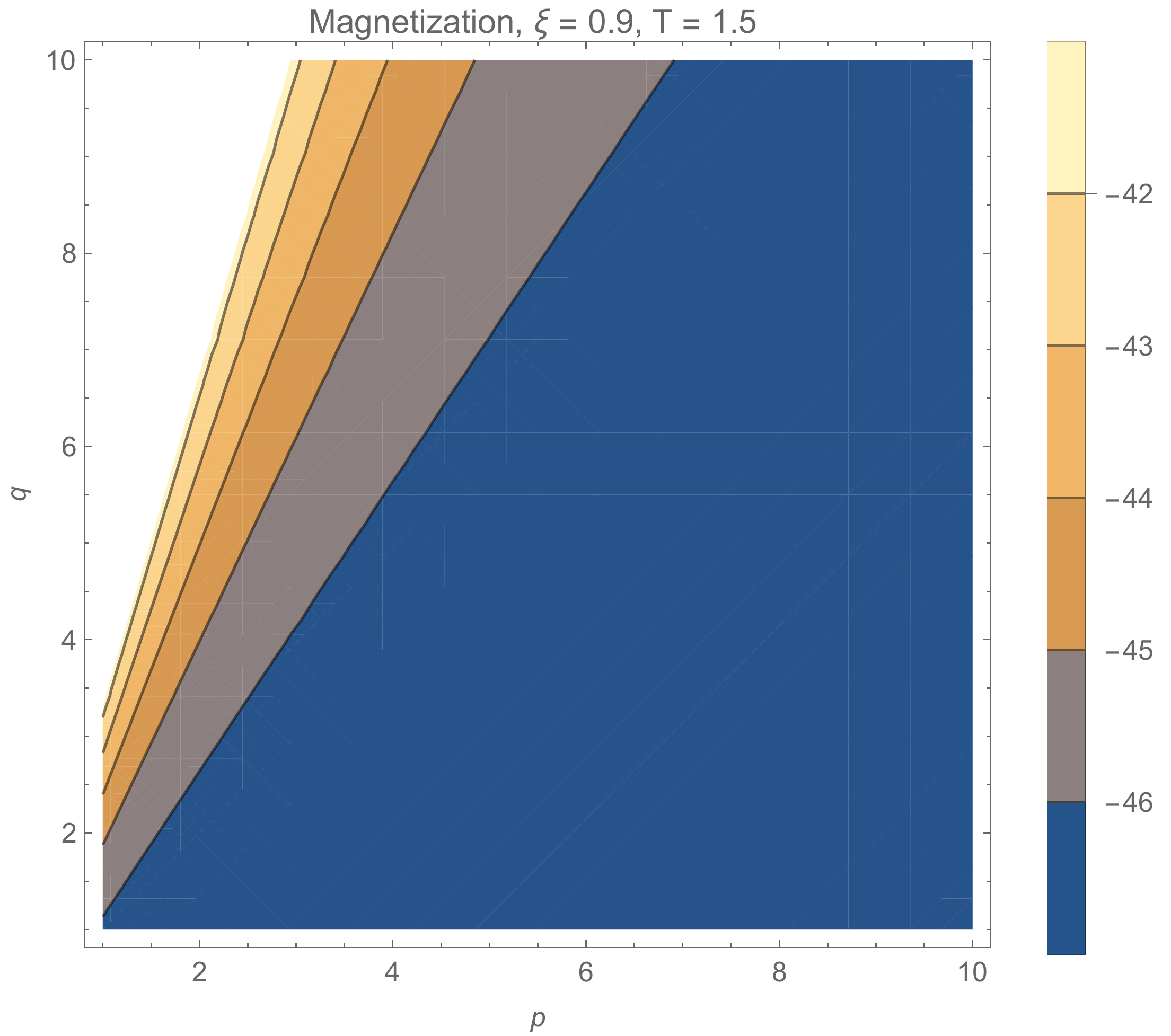}
\includegraphics[scale=0.35]{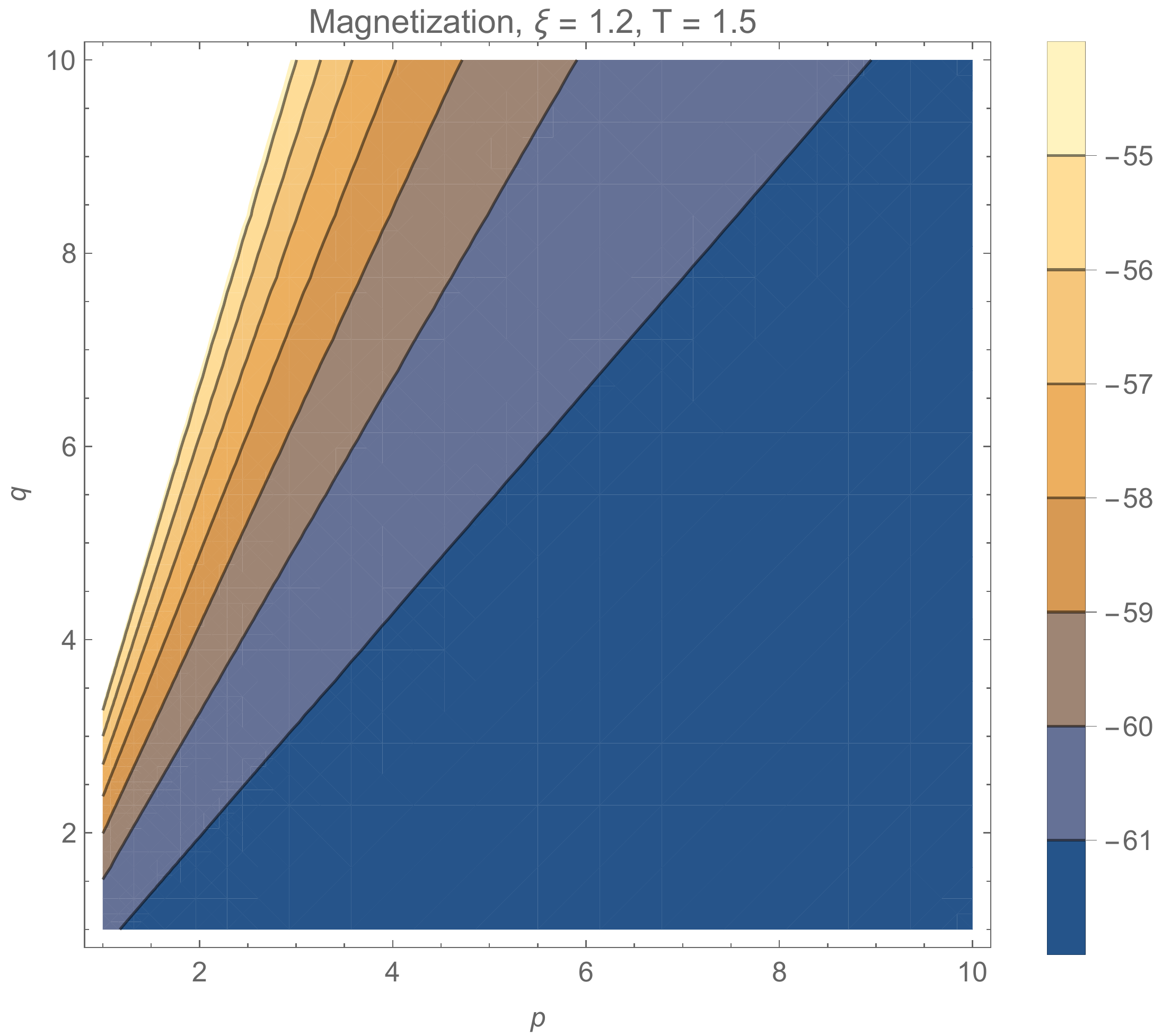}\\
\includegraphics[scale=0.35]{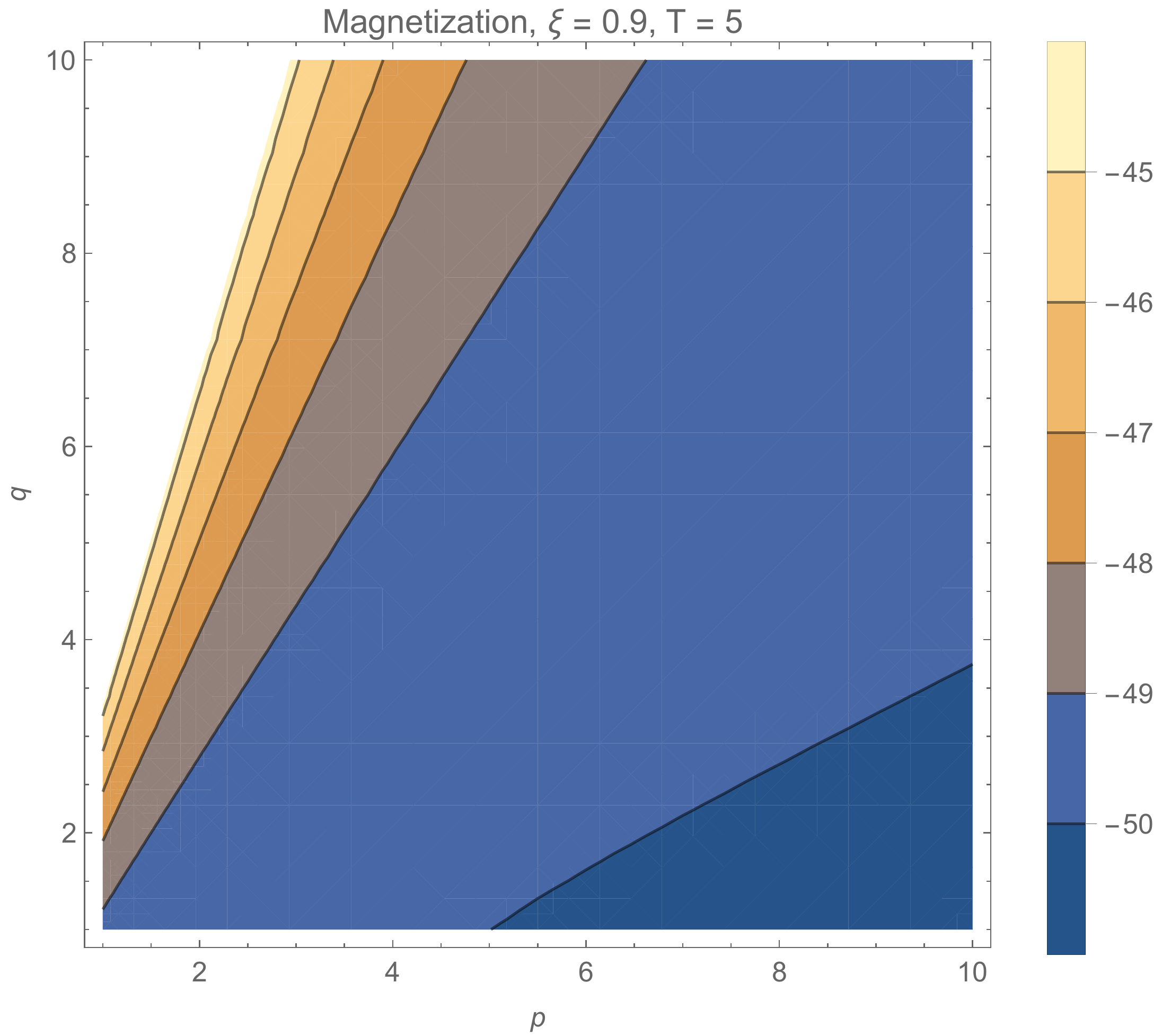}
\includegraphics[scale=0.35]{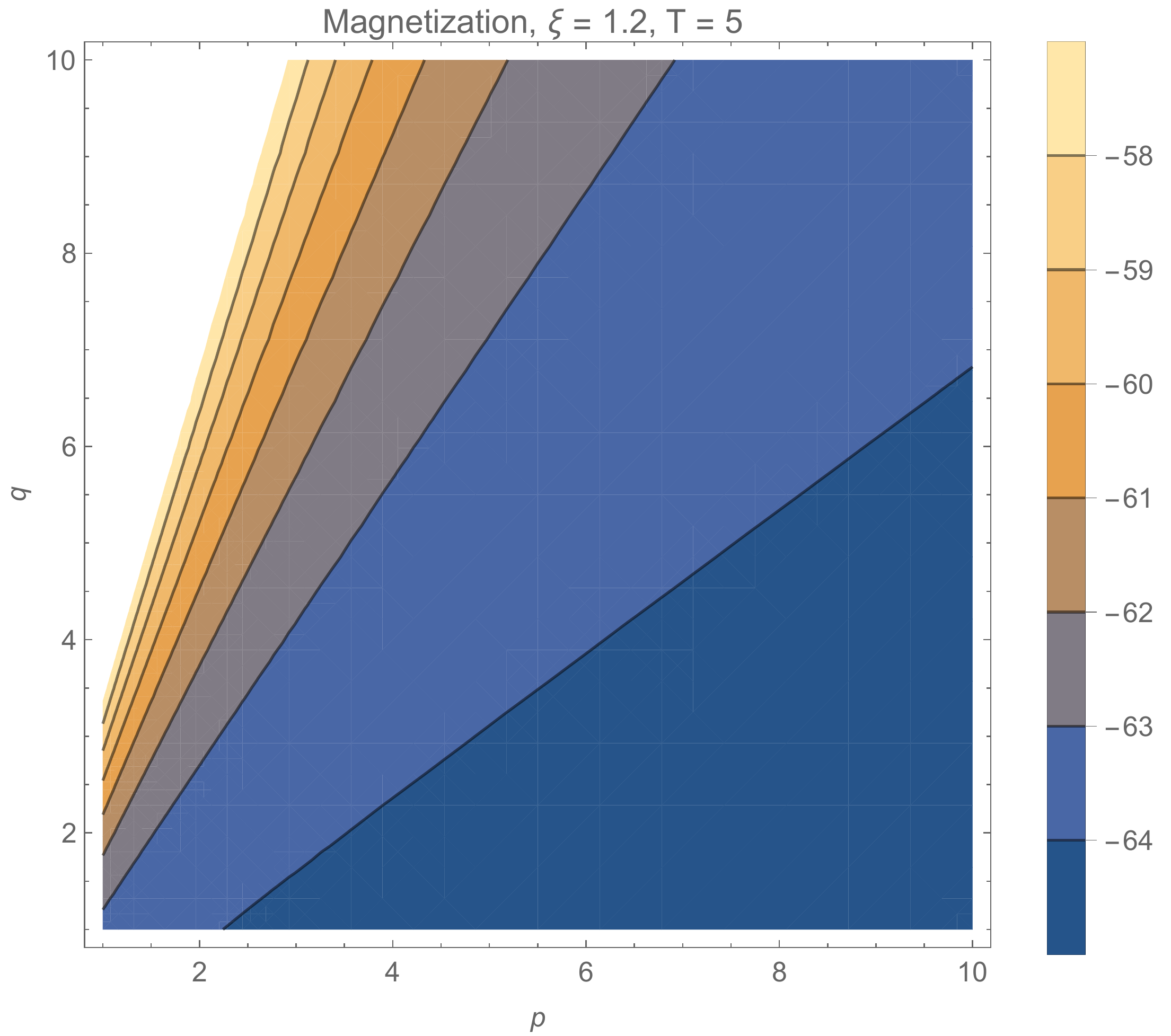}
\caption{These contour plots show the magnetization for different configurations of temperature and magnetic field. In these ones, we can see the behavior of the magnetization as a function of the wind numbers $(p,q)$.}
\label{fig:Magnetization3D}
\end{figure}

\begin{figure}[h!]
\centering
\includegraphics[scale=0.35]{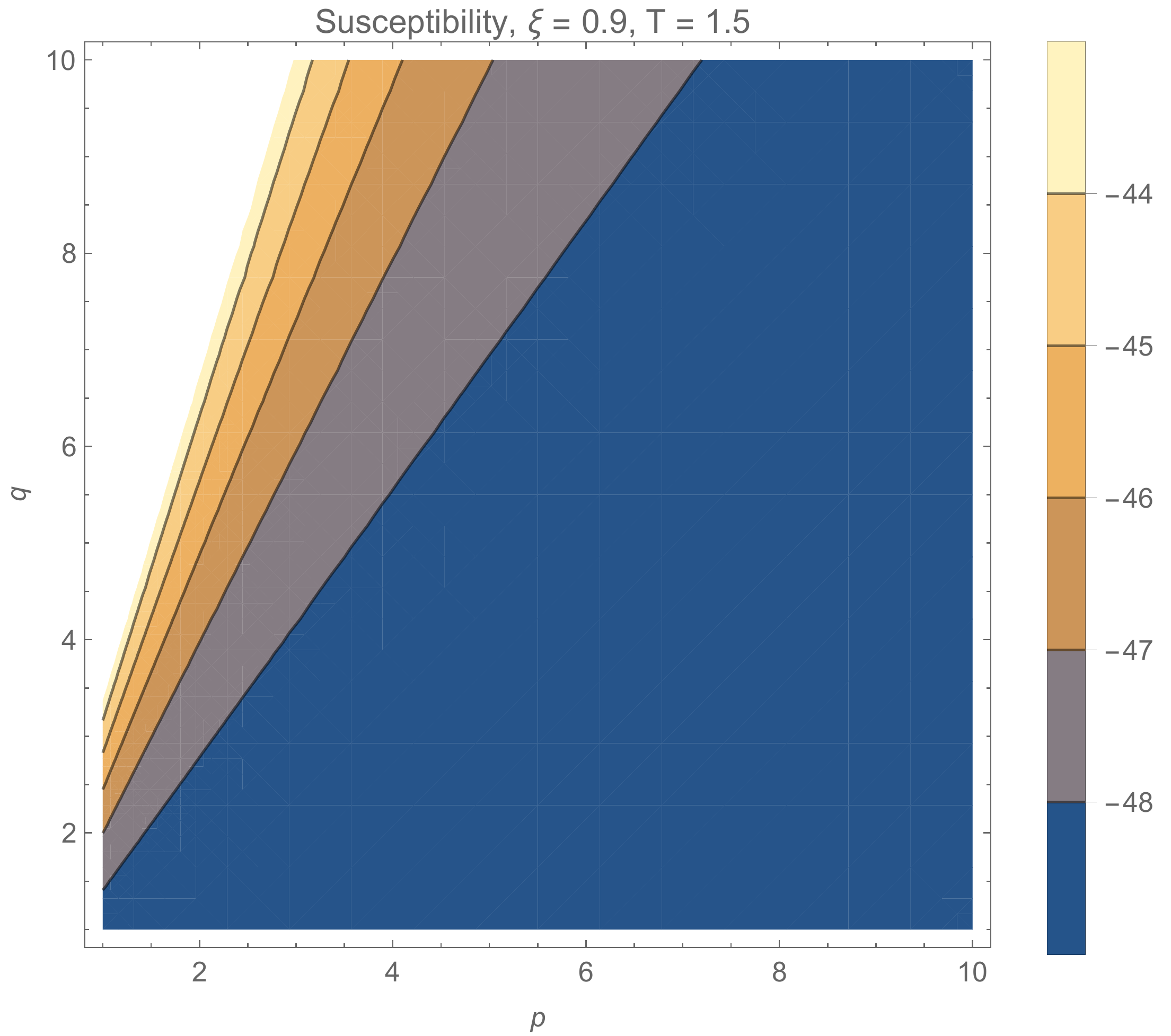}
\includegraphics[scale=0.35]{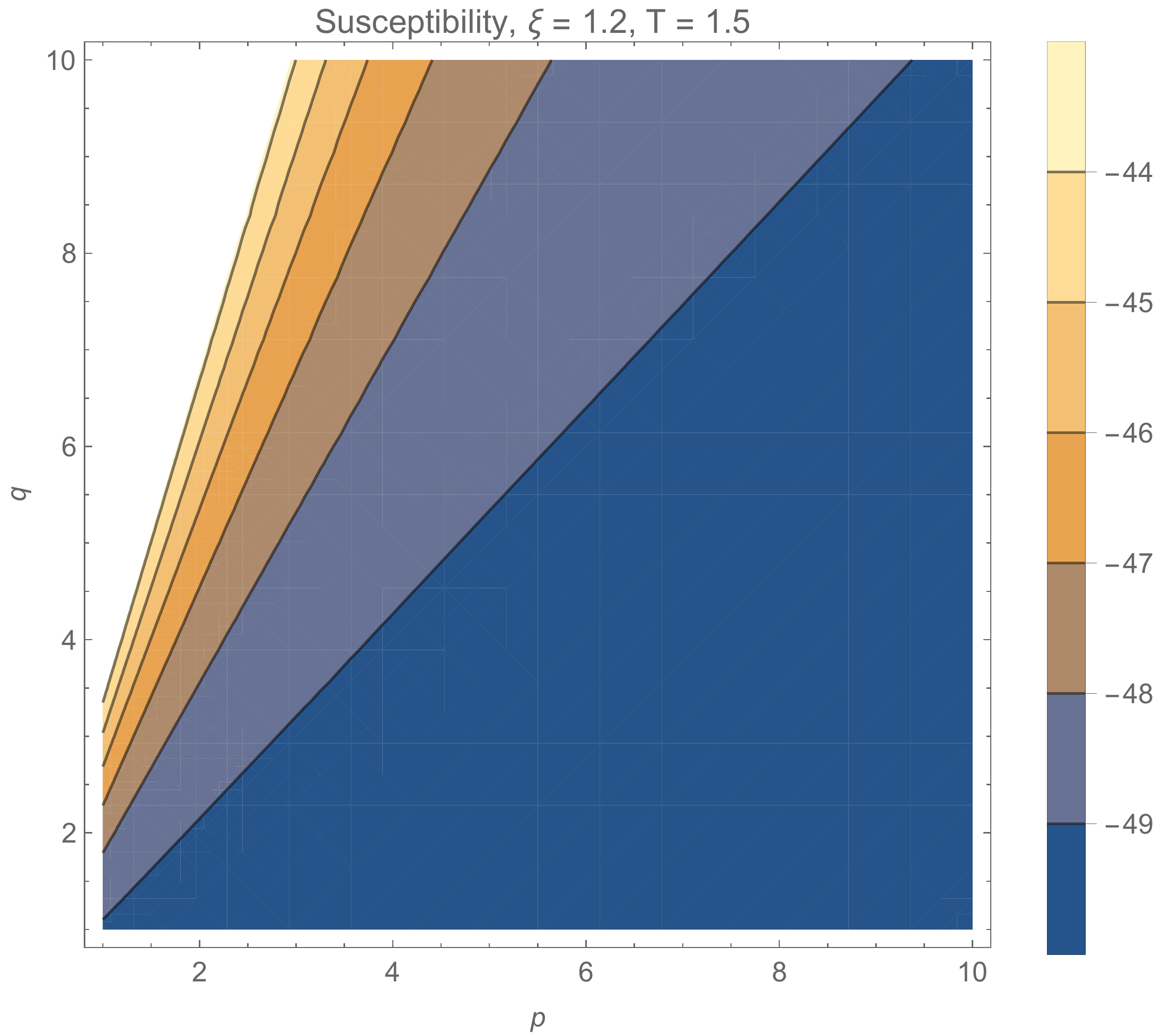}\\
\includegraphics[scale=0.35]{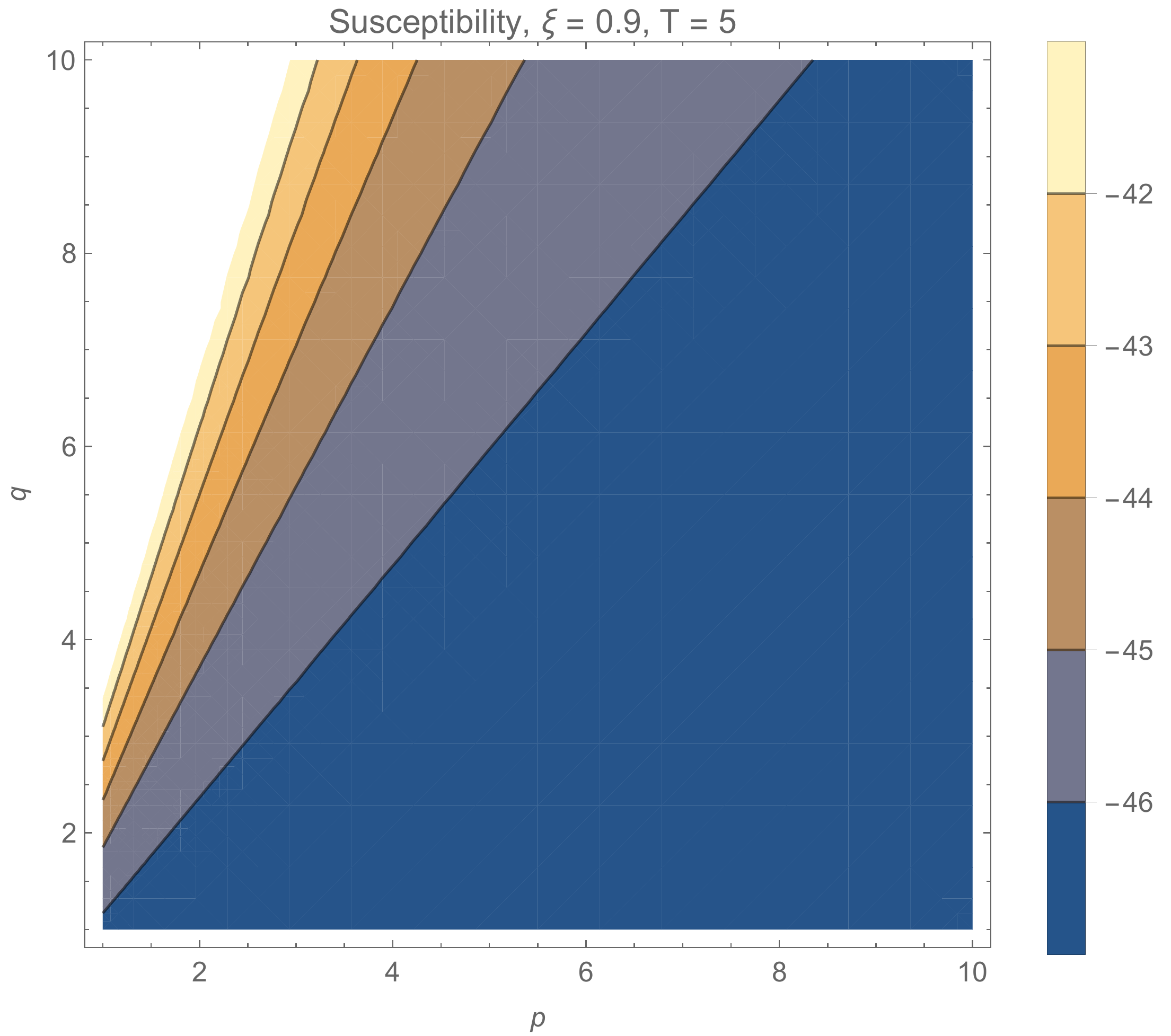}
\includegraphics[scale=0.35]{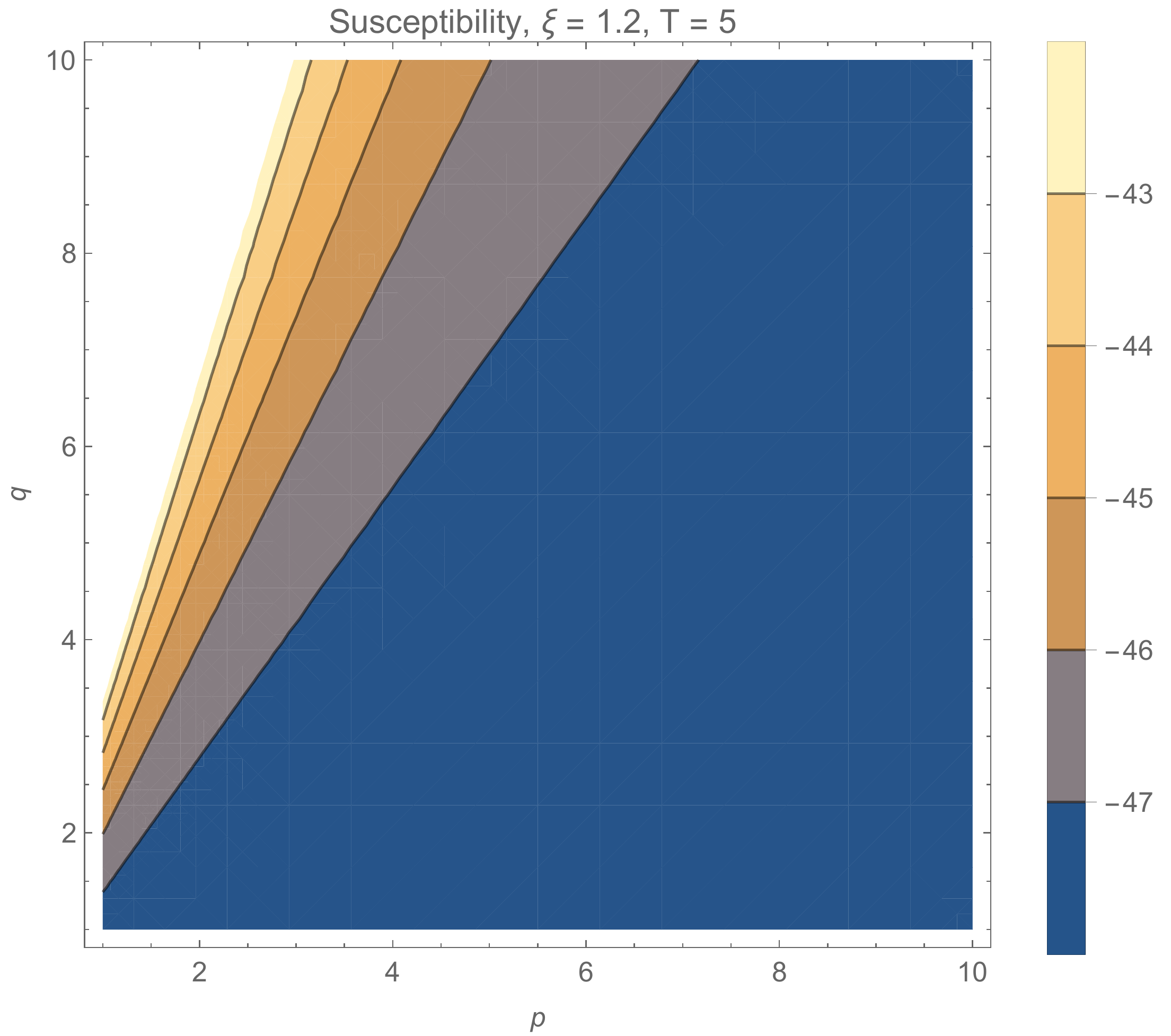}
\caption{The contour plots show how the susceptibility varies for different configurations of temperature and magnetic field. In these plots, we see the behavior of the Susceptibility as a function of the wind numbers $(p,q)$.}
\label{fig:Susceptibility3D}
\end{figure}

In order to acquire a better understanding of the behavior of the thermodynamic functions for different combinations of toroidal and poloidal windings  $p$ and $q$, we display below the contour plot of such functions with respect to certain ranges of  winding numbers, as displayed in Figs. \ref{fig:Helmholtz3D}, \ref{fig:Entropy3D}, \ref{fig:Energy3D}, \ref{fig:HCapacity3D}, \ref{fig:Magnetization3D}, and \ref{fig:Susceptibility3D}. Unlike the special cases of zero magnetic field and/or zero temperature, in general situations, the thermodynamic observables depend on both $p,q$ individually (i.e.  not only on $\alpha$).  

The plots mainly compare how those functions behave for a given temperature and magnetic field. Initially, in Fig. \ref{fig:Helmholtz3D}, we see the Helmholtz free energy for different fixed values of temperature and magnetic field -- the variable combinations of  winding numbers $(p,q)$, including  co-prime $p,q$, satisfy the knot condition. We see in all configurations that the larger  the values of $(p,q)$, the larger the Helmholtz energy becomes. In Fig. \ref{fig:Entropy3D}, we also investigate how entropy is modified for different configurations of temperature and magnetic field. It is shown the behavior of the entropy as a function of the winding numbers $(p,q)$. In short, we verify that in all configurations when we increase  values of $(p,q)$, entropy increases. Next, in Fig. \ref{fig:Energy3D}, we provide the contour plot for the mean energy regarding different configurations of temperature and magnetic field as well. We see that in all configurations when  $q$ is increased for fixed   $p$, the mean energy decreases; on the other hand, when the values of $p$ increase maintaining $q$ fixed, the opposite behavior occurs.  Fig. \ref{fig:HCapacity3D} displays contour plots to the heat capacity for different values of temperature and magnetic field. In these ones, we can naturally see the behavior of the heat capacity as a function of the winding numbers $(p,q)$. We see that in all configurations when we increase the values of $q$ keeping $p$ fixed, the heat capacity decreases its values; on the other hand, if we have the values of $p$ increased maintaining $q$ fixed, the opposite behavior also occurs. In Fig. \ref{fig:Magnetization3D}, we also show the plots exhibiting how the magnetization is affected by different configurations of temperature and magnetic field. Here, we realize the behavior of the magnetization as a function of the winding numbers $(p,q)$. Thereby, we see that in all configurations when we increase the values of $(p,q)$, the larger the magnetization becomes. Finally, in Fig. \ref{fig:Susceptibility3D}, the plots exhibit how the the susceptibility is changed for different temperature and magnetic field. In this sense, we see the behavior of the susceptibility as a function of the winding numbers $(p,q)$. Furthermore,  we verify that in all configurations when we increase the values of $(p,q)$,    susceptibility becomes larger.

\subsection{Thin torus: the limit $a\rightarrow0$}
In this subsection, we focus on a particular case when the limit $a\rightarrow 0$ is taken into account. In other words, this means that we have a \textquotedblleft
circle\textquotedblright\ or a thin torus. Thereby, we get the following single
particle partition function
\begin{equation}
\mathcal{Z}_{1}=\sqrt{\frac{\pi mp^{2}d^{2}}{2\beta }}\left\{ 1+\mathrm{erf}%
\left[ -\frac{\xi }{4}\sqrt{\frac{2\beta }{m\left( 1\right) }}\right]
\right\} . \label{eq:Exact_Z0}
\end{equation}
Using the same methodology, we can also get from the above result the Helmholtz free energy
\begin{equation}
f=-\frac{1}{\beta }\ln \sqrt{\frac{\pi mp^{2}d^{2}}{2\beta }}-\frac{1}{\beta
}\ln \left\{ 1+\mathrm{erf}\left[ -\frac{\xi }{4}\sqrt{\frac{2\beta }{m}}%
\right] \right\},
\end{equation}%
and the internal energy per particle
\begin{eqnarray}
u &=&-\frac{1}{\beta ^{2}}\left( \frac{1}{2}+\ln \sqrt{\frac{\pi mp^{2}d^{2}%
}{2\beta }}\right) -\frac{1}{\beta ^{2}}\ln \left\{ 1+\mathrm{erf}\left( -%
\frac{\xi }{4}\sqrt{\frac{2\beta }{m}}\right) \right\} \notag \\ 
&&\phantom{ss}-\frac{\xi }{\beta ^{\frac{3}{2}}}\sqrt{\frac{1}{8m\pi }}\frac{%
\exp \left( -\frac{\xi ^{2}}{8}\frac{\beta }{m}\right) }{1-\mathrm{erf}%
\left( \frac{\xi }{4}\sqrt{\frac{2\beta }{m}}\right) }.
\end{eqnarray}

The magnetization and susceptibility, respectively, are displayed below:
\begin{equation}
\mathfrak{m}=-\sqrt{\frac{1}{2\pi \beta m}}\frac{\exp \left( -\frac{\xi ^{2}%
}{8}\frac{\beta }{m}\right) }{1-\mathrm{erf}\left( \frac{\xi }{4}\sqrt{\frac{%
2\beta }{m}}\right) },
\end{equation}%
\begin{equation}
\chi =-\frac{4}{\pi \beta }\frac{\mathcal{B}^{2}\exp \left( -2\mathcal{B}%
^{2}\xi ^{2}\right) }{1-\mathrm{erf}\left( \mathcal{B}\xi \right) }\left\{
\frac{1}{1-\mathrm{erf}\left( \mathcal{B}\xi \right) }-\sqrt{\pi }\mathcal{B}%
\xi \exp \left( \mathcal{B}^{2}\xi ^{2}\right) \right\} ,
\end{equation}%
where the parameter $\mathcal{B}$ assumes the form: 
\begin{equation}
\mathcal{B}=\frac{1}{4}\sqrt{\frac{2\beta }{m}}.
\end{equation}%

Let us now calculate the magnetization and the susceptibility for the same particular configuration of magnetic field and temperature as we did before. Initially, we
consider a configuration in which $\xi =0$. For this case, we obtain the
following results:
\begin{equation}
\mathfrak{m}_{\xi =0} =-\frac{1}{\sqrt{2\pi \beta m}}, \phantom{aa}\chi _{\xi =0} =-\frac{1}{2 \pi m }.
\end{equation}%
Now, for both $\xi =0$ and $T=0$, we get%
\begin{equation}
\mathfrak{m}_{\xi =0,T=0} =0,\phantom{a}\chi _{\xi =0,T=0} =-\frac{1}{2\pi m}.
\end{equation}

Here we notice that for $\xi =0$ and $T=0$, the magnetization turns out to be zero and we have a non-null constant susceptibility. On the other hand, considering the configuration where $\xi =0$, both magnetization and susceptibility are non-zero. We have to pay attention in the fact that the magnetization (when $\xi =0$) has the dependence on $T^{\frac{1}{2}}$. Another feature that is worth noticing is the fact that the mass ascribed to the fermions under consideration determines the magnitude of the susceptibility at zero magnetic field and also at zero temperature.

\section{Fermions on a torus knot}\label{Sec:Fermions}

We intend now to apply the grand canonical ensemble approach to understand the behavior of $N$ noninteracting electrons moving in a prescribed  torus knot on  a single torus. Since this formalism  takes into account the Fermi-Dirac statistics, we can obtain more information on how the Pauli principle can modify the properties of the system.

The grand partition function for the present
problem reads%
\begin{equation}
\Xi =\sum_{N=0}^{\infty }\exp \left( \beta \mu N\right) \mathcal{Z}\left[
N_{\Omega }\right] ,  \label{eq:GarndPartition-function}
\end{equation}%
where $\mathcal{Z}\left[ N_{\Omega }\right] $ is the canonical partition
function which is now a function of the occupation number $N_{\Omega }$ and $%
\Omega $ label a quantum state. Since we are dealing with fermions, we know
that the occupation number allowed for each quantum state is restricted to $%
N_{\Omega }=\left\{ 0,1\right\} $. So, for an arbitrary quantum state, the
energy depends also on the occupation number as
\begin{equation*}
E\left\{ N_{\Omega }\right\} =\sum_{\left\{ \Omega \right\} }N_{\Omega
}E_{\Omega }
\end{equation*}%
where we have the total fermion number $N$ defined as%
\begin{equation*}
\sum_{\left\{ \Omega \right\} }N_{\Omega }=N.
\end{equation*}%
Thus the partition function becomes%
\begin{equation}
\mathcal{Z}\left[ N_{\Omega }\right] =\sum_{\left\{ N_{\Omega }\right\}
}\exp \left[ -\beta \sum_{\left\{ \Omega \right\} }N_{\Omega }E_{\Omega }%
\right] .
\end{equation}%
The grand partition function assumes the form%
\begin{equation}
\Xi =\sum_{N=0}^{\infty }\exp \left( \beta \mu N\right) \sum_{\left\{
N_{\Omega }\right\} }\exp \left[ -\beta \sum_{\left\{ \Omega \right\}
}N_{\Omega }E_{\Omega }\right] ,
\end{equation}%
which can be rewritten as
\begin{equation}
\Xi =\prod_{\left\{ \Omega \right\} }\left\{ \sum_{\left\{ N_{\Omega
}\right\} }\exp \left[ -\beta N_{\Omega }\left( E_{\Omega }-\mu \right) %
\right] \right\} .
\end{equation}%
Performing the sum over the possible occupation numbers, we obtain%
\begin{equation}
\Xi =\prod_{\left\{ \Omega \right\} }\left\{ 1+\exp \left[ -\beta \left(
E_{\Omega }-\mu \right) \right] \right\} ,
\end{equation}

The connection with thermodynamics is made using the grand potential given by%
\begin{equation}
\Phi =-\frac{1}{\beta }\ln \Xi .
\end{equation}%
Replacing $\Xi $ in the equation above, we get%
\begin{equation}
\Phi =-\frac{1}{\beta }\sum_{\left\{ \Omega \right\} }\ln \left\{ 1+\exp %
\left[ -\beta \left( E_{\Omega }-\mu \right) \right] \right\} .
\label{eq:Gand-potential}
\end{equation}%
The entropy of the system can be cast in the following compact form, namely%
\begin{equation*}
S=-\frac{\partial \Phi }{\partial T}=-k_{B}\sum_{\left\{ \Omega \right\} }%
\mathcal{N}_{\Omega }\ln \mathcal{N}_{\Omega }+\left( 1-\mathcal{N}%
_{\Omega }\right) \ln \left( 1- \mathcal{N}_{\Omega }\right)
\end{equation*}%
where we explicitly use the Fermi-Dirac distribution function%
\begin{equation*}
\mathcal{N}_{\Omega }=\frac{1}{\exp \left[ \beta \left( E_{\Omega }-\mu
\right) \right] +1}.
\end{equation*}%
We can also use the grand potential to calculate other important
thermodynamics properties as mean particle number, energy, heat capacity and
pressure using the following equation, respectively,%
\begin{eqnarray}
\mathcal{N} =-\frac{\partial \Phi }{\partial \mu },\phantom{ss} \mathcal{U} =-T^{2}\frac{\partial }{\partial T}\left( \frac{\Phi }{T}\right),\phantom{ss} C_{V} =T\frac{\partial S}{\partial T}.
\end{eqnarray}%
we can also calculate, as before, the magnetization and the susceptibility.

In order to calculate all thermodynamics quantities described above, we need to perform before the sum present in Eq. $\left( \ref{eq:Gand-potential}%
\right) $. Fortunately, it is possible to do it in a closed form using again
the \textit{Euler-MacLaurin }formula. Before performing the sum, we will
rewrite the energy $\left( \ref{q:SEnergy}\right) $ in a more convenient
form%
\begin{equation}
E_{n}=\frac{1}{2m}\frac{\hbar ^{2}\left( 1-\mathcal{A}\right) }{p^{2}d^{2}}%
\left[ n+\frac{\xi }{\hbar }\frac{pd^{2}}{\left( 1-\mathcal{A}\right) }%
\left( 1-\mathcal{A}+\frac{\alpha^{2}}{2}\mathcal{A}\right) \right] ^{2}-\frac{\xi ^{2}%
\alpha^{4}\mathcal{A}^{2}d^{2}}{32m\left( 1-\mathcal{A}\right) }.
\end{equation}%
Thereby, the grand partition function $\left( \ref{eq:Gand-potential}\right) \,\ $can
be rewritten as
\begin{equation}
\Phi =-\frac{2}{\beta }\int_{0}^{\infty } \mathrm{d}E \,\ln \left\{ 1+z\exp \left[ -\beta
E\left( n\right) \right] \right\} dn-\frac{1}{\beta }\ln \left\{
1+ze^{-\beta \Upsilon }\right\} ,
\end{equation}%
where%
\begin{equation*}
\Upsilon \equiv \frac{\xi ^{2}}{2m}\frac{d^{2}}{\left( 1-%
\mathcal{A}\right) }\left( 1-\mathcal{A}+\frac{\alpha^{2}}{2}\mathcal{A}\right) ^{2}-%
\frac{\xi ^{2}\alpha^{4}\mathcal{A}^{2}d^{2}}{32m\left( 1-\mathcal{A}\right) }
\end{equation*}%
Performing now the integration, we get%

\begin{equation}
\Phi =\frac{\mathcal{L}}{\lambda }f_{\frac{3}{2}}\left( \Lambda ,\mathfrak{X}%
\right) -f_{1}\left( \mathfrak{X}\right) .  \label{eq:GCanonicalFermion}
\end{equation}%
where $f_{\sigma }\left( \Lambda ,\mathfrak{X}%
\right) $

\begin{equation}
f_{\sigma }\left( \Lambda ,\mathfrak{X}\right) =\frac{1}{\Gamma \left(
\sigma \right) }\int_{\Lambda }^{\infty }\frac{t^{\sigma -1}}{\mathfrak{X}%
^{-1}e^{t}+1}\mathrm{d}t,
\end{equation}%
is the incomplete Fermi-Dirac integral. We also define the quantities%
\begin{eqnarray}
\mathfrak{X} &=&ze^{\beta \Psi }, \\
\Psi &=&\frac{\xi ^{2}\alpha^{4}\mathcal{A}^{2}d^{2}}{32m\left( 1-\mathcal{A}\right) },
\\
\Lambda &=&\frac{\xi }{\hbar p}\frac{p^{2}d^{2}}{\left( 1-\mathcal{A}\right)
}\left( 1-\mathcal{A}+\frac{\alpha^{2}}{2}\mathcal{A}\right) , \\
\mathcal{L} &=&\frac{p^{2}d^{2}}{\left( 1-\mathcal{A}\right) }, \\
\lambda &=&\sqrt{\frac{2\pi \hbar ^{2}}{m}\beta }.
\end{eqnarray}%
For instance, the entropy is given by%
\begin{equation}
S=\frac{\pi k_{B}\hbar ^{2}}{m}\frac{\mathcal{L}}{\lambda ^{3}}f_{\frac{3}{2}%
}\left( \Lambda ,\mathfrak{X}\right) +k_{B}\left( \Psi +\mu \right) \frac{%
\mathcal{L}}{\lambda }f_{\frac{1}{2}}\left( \Lambda ,\mathfrak{X}\right) -%
\frac{k_{B}\left( \Psi +\mu \right) ze^{-\beta \Upsilon }}{1+ze^{-\beta
\Upsilon }}.
\end{equation}

\begin{figure}[tbh]
\centering
\includegraphics[width=8cm,height=5cm]{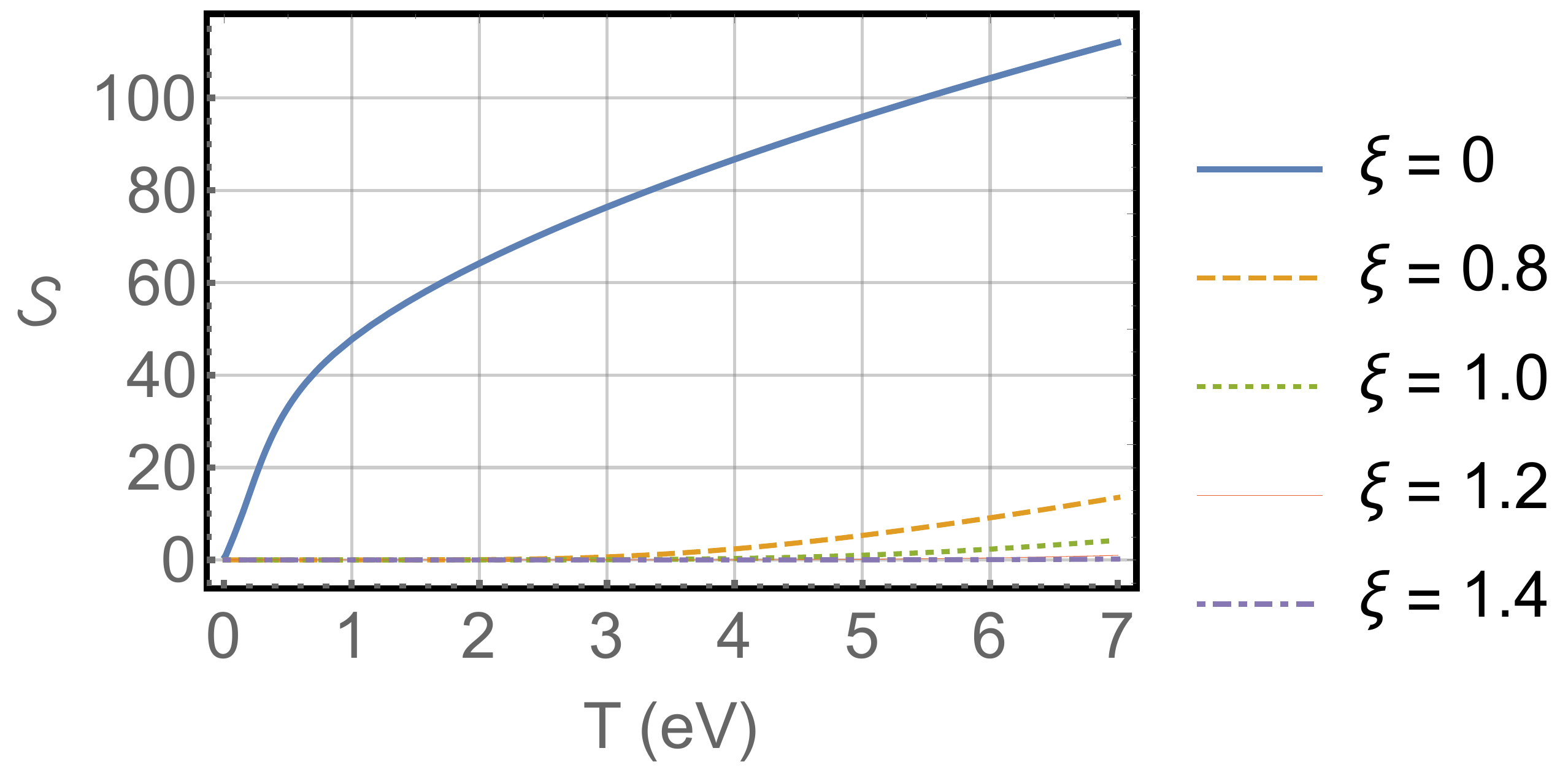}
\includegraphics[width=8cm,height=5cm]{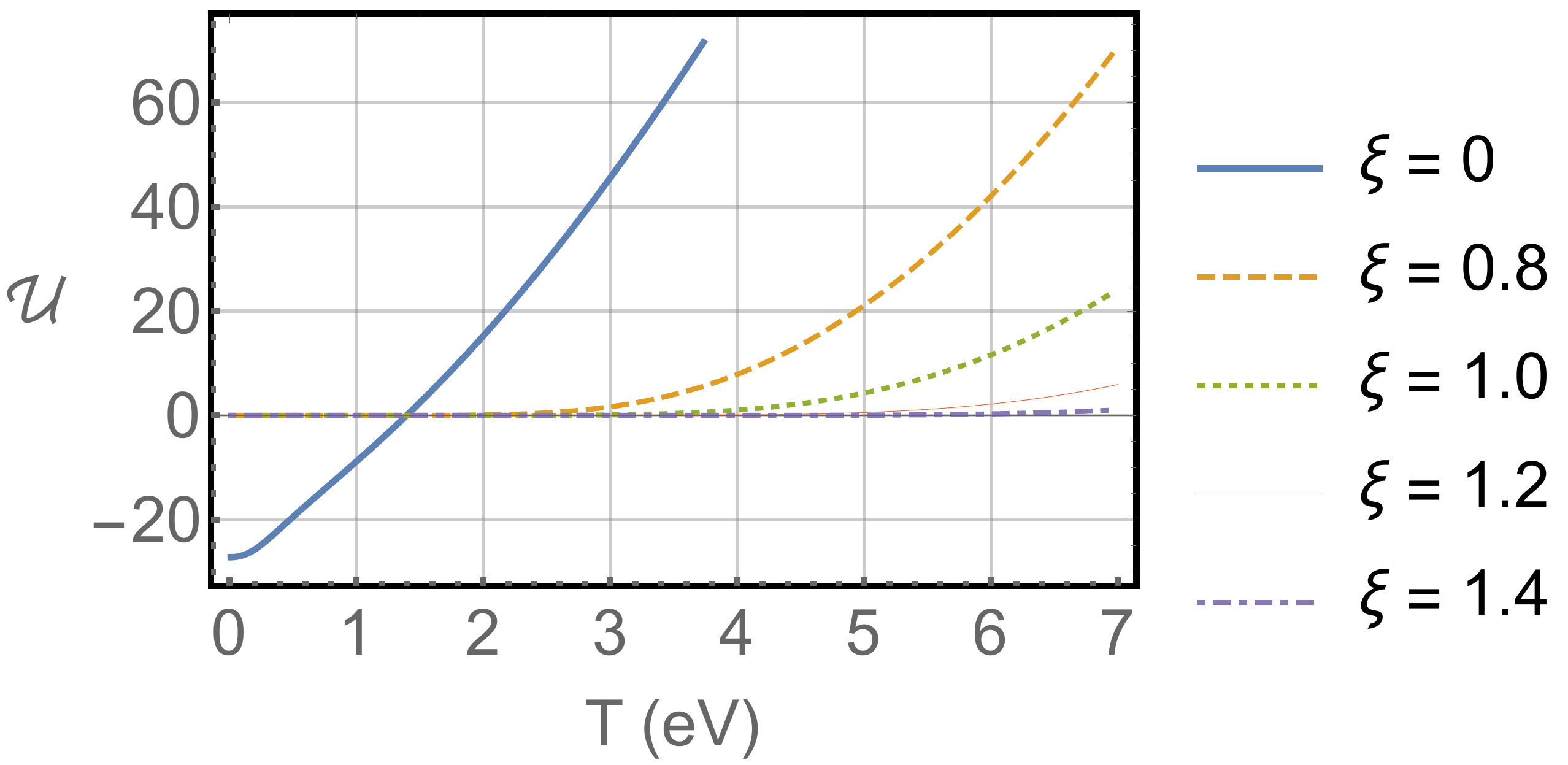}
\caption{Entropy and internal energy}
\label{Fig:Entropy2}
\end{figure}

\begin{figure}[tbh]
\centering
\includegraphics[width=8cm,height=5cm]{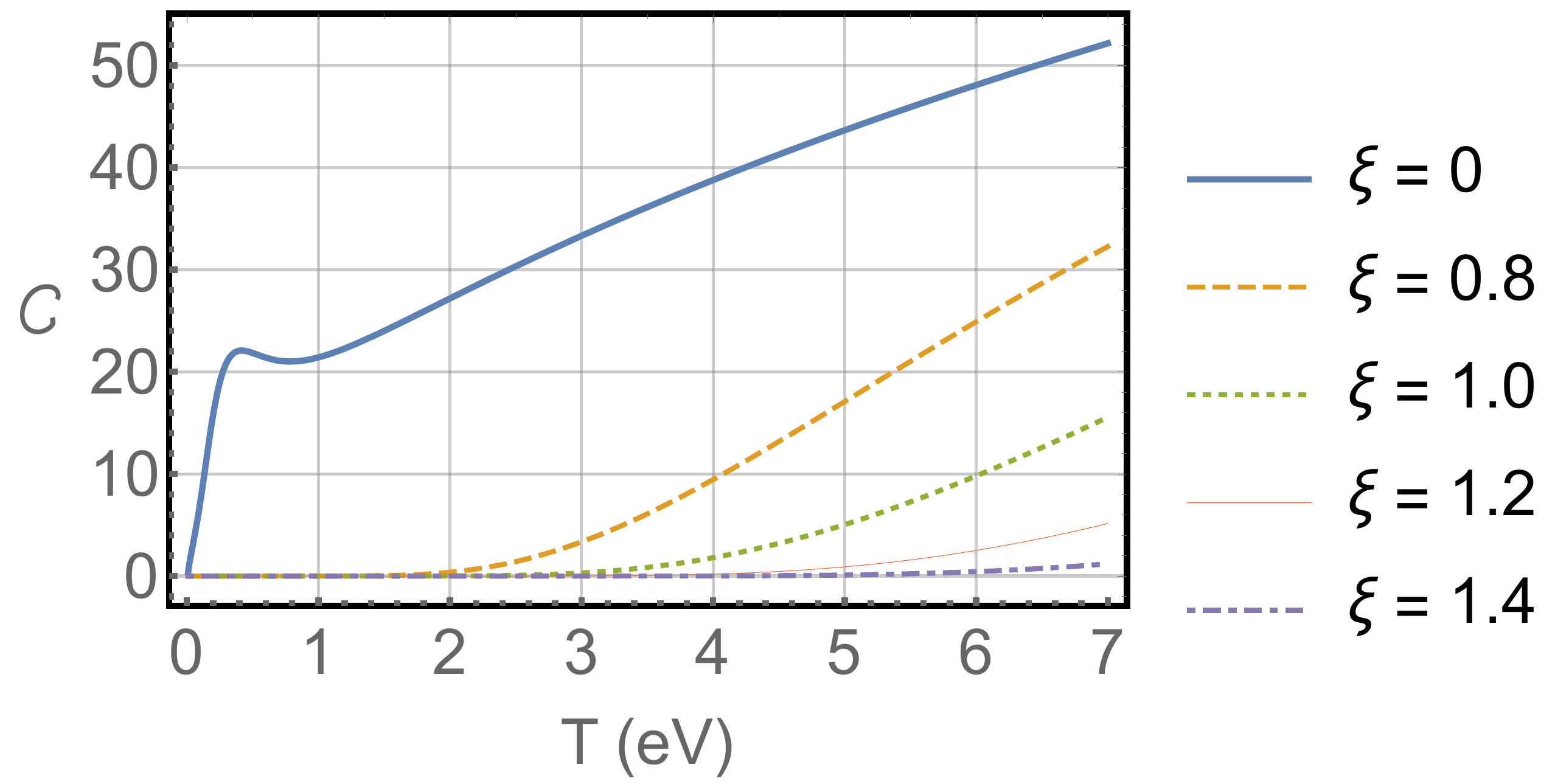}
\caption{Heat capacity}
\label{Fig:Hcapacity2}
\end{figure}

\begin{figure}[tbh]
\centering
\includegraphics[width=8cm,height=5cm]{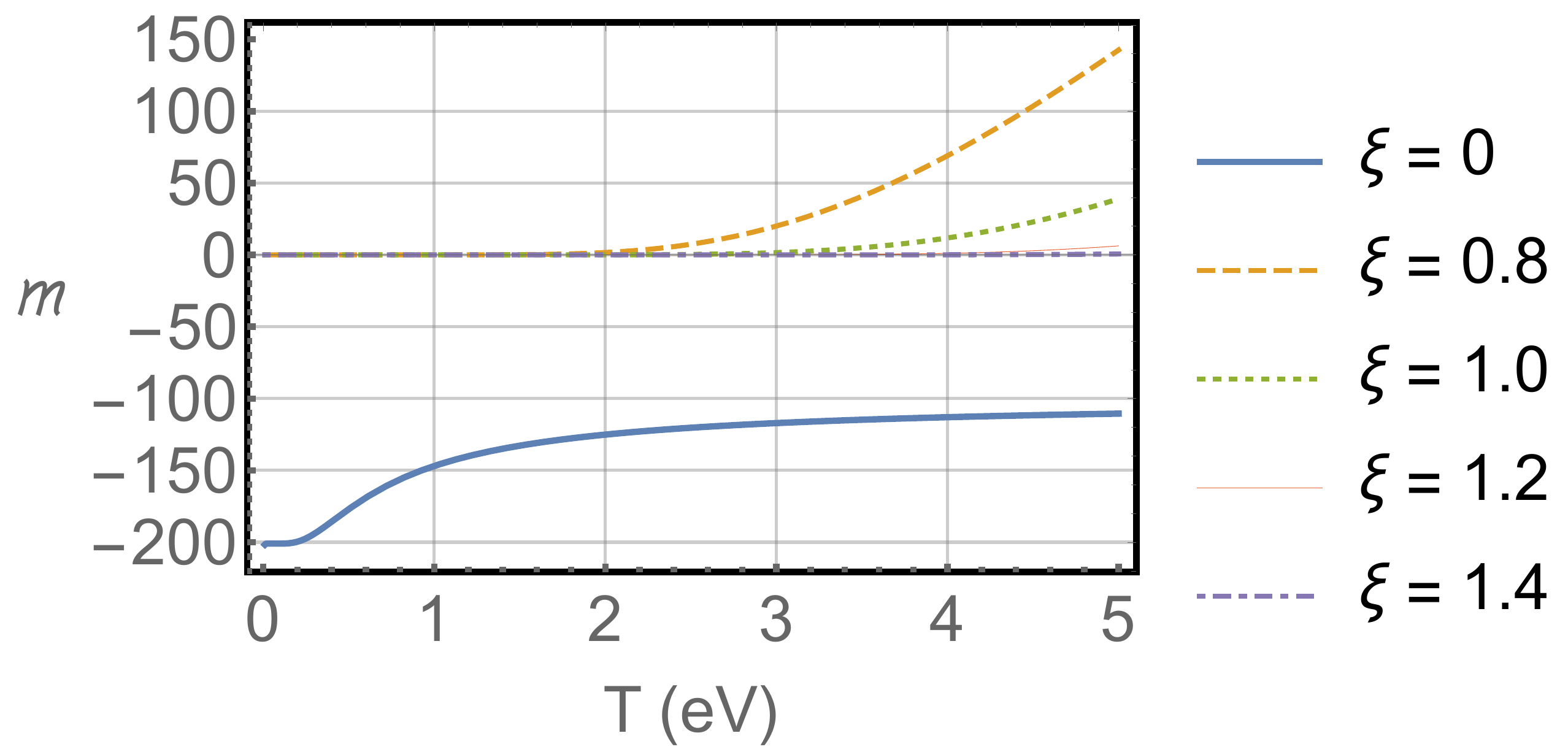}
\includegraphics[width=8cm,height=5cm]{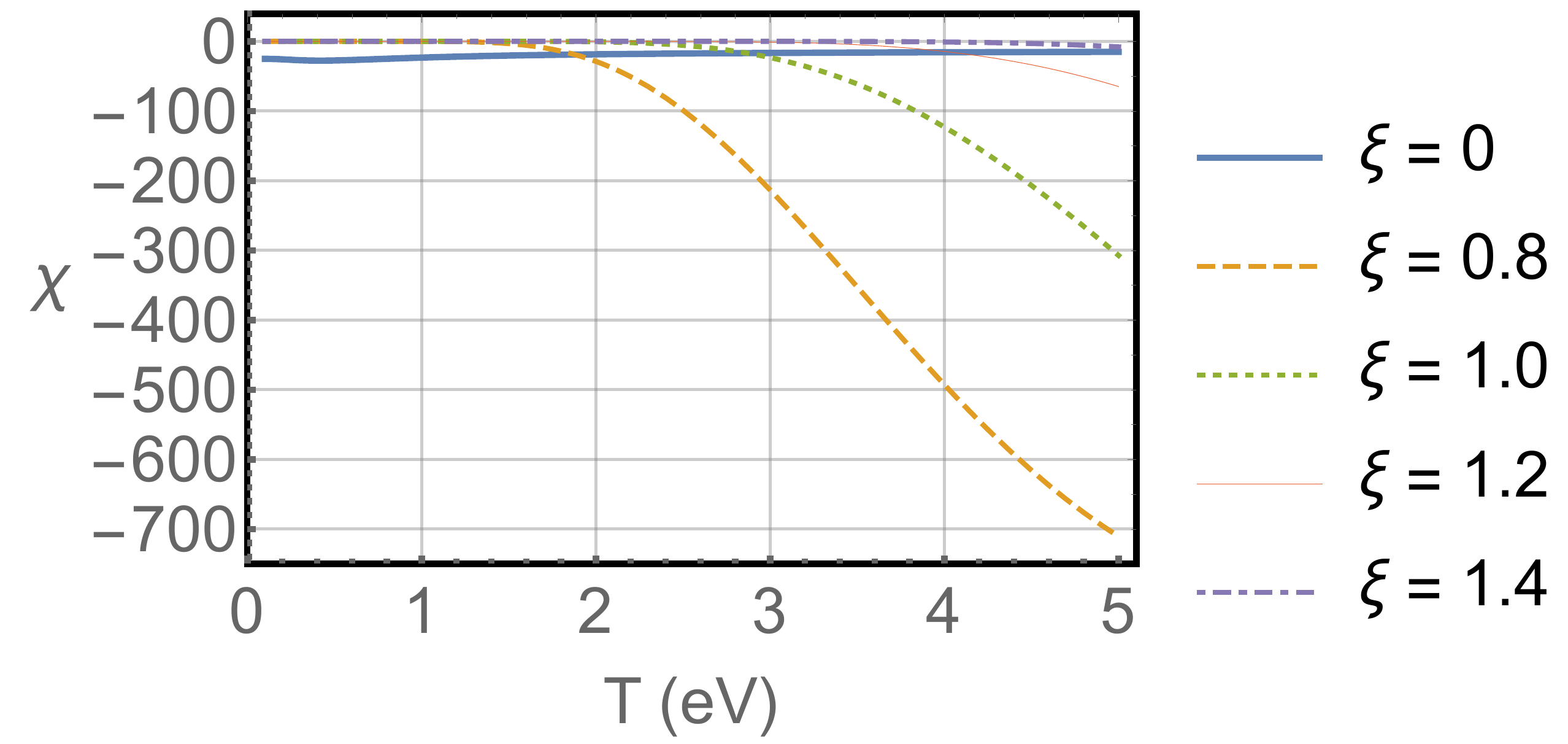}
\caption{Magnetization and susceptibility}
\label{Fig:Magnetization2}
\end{figure}

In order to give an idea of the behavior of the thermodynamic functions, we display below, in Fig. \ref{Fig:Entropy2}, \ref{Fig:Hcapacity2} and \ref{Fig:Magnetization2}, some plots taking into account the same values as those presented in Sec. \ref{Sec:N-torus}. Here, we see that for entropy, mean energy, heat capacity, and magnetization, when the temperature rises, the magnitude of their thermal properties increases; on the other hand, to the susceptibility the opposite behavior happened: when the temperature raised, the magnitude of such thermodynamic function decreased. On the other hand, when the magnetic field increases, the magnitude of entropy, mean energy and heat capacity decrease their values for a fixed temperature. However, the magnetization has an opposite behavior.

As an application, let us use the result obtained from the Eq. $\left( \ref%
{eq:GCanonicalFermion}\right) $ to probe how interaction affects the Fermi
energy. From Eq. $\left( \ref%
{eq:GCanonicalFermion}\right) $ we get

\begin{equation}
N=g\left[ \frac{\mathcal{L}}{\lambda }f_{\frac{1}{2}}\left( \Lambda ,%
\mathfrak{X}\right) -\frac{\mathfrak{X}}{\mathfrak{X}+1}\right] ,
\end{equation}%
where $g$ is a weight factor that arises from the internal structure of the
particles. The Fermi energy $\mu _{0}$ is the energy of the topmost filled
level in the ground state of the $N$ electron system. In this way, we get%
\begin{equation}
N=g\left\{ \frac{4\mathcal{L}}{3}\left[ \frac{m}{2\pi ^{2}\hbar ^{2}}\left(
\Psi +\mu _{0}\right) \right] ^{\frac{1}{2}}-1\right\} .
\end{equation}%
Solving the above equation for $\mu _{0}$, we get%
\begin{equation}
\mu _{0}=\frac{9\pi ^{2}\hbar ^{2}}{8m\mathcal{L}^{2}}\left( \frac{N}{g}%
+1\right) ^{2}-\Psi .
\label{eq:fermi-energy-nonI}
\end{equation}

It is interesting to notice that the Fermi energy $\mu_0$ consists of two clearly divided parts. The first one contains the all information of a generic system; and all features of the specific torus (and torus-knot path) are encountered in the second part, $\Psi$. When we reduce the torus to a ring -- when the limit $a \gg d$ is taken into account, we recover the well known result established in the literature \cite{dai2003quantum,dai2004geometry}.

\section{Interacting Fermions on a Torus Knot}\label{Sec:Interacting-Fermions}

Following  the interacting approach developed in  [54], we can derive the
modified grand canonical potential taking into account the interaction
between fermions, namely%
\begin{eqnarray}
\Phi  &=&-T\ln \mathcal{Z}  \notag \\
&=&-T\sum_{\Omega }\ln \left( 1+\exp \left[ -\beta \left( E_{\Omega
}+u^{\prime }\left( \bar{n}\right) -\mu \right) \right] \right) +U\left( V,%
\bar{n}\right) -u^{\prime }\left( \bar{n}\right) \bar{N}.
\label{eq:grand-canonical-potencial}
\end{eqnarray}%
The interaction is incorporated through $u^{\prime }\left( \bar{n}\right)$ term.
Based on this equation, the other thermodynamic functions can be calculated
as well. Performing as before the \textit{Euler-MacLaurin }formula we find%
\begin{equation}
\Phi =\frac{\mathcal{L}}{\lambda }f_{\frac{3}{2}}\left( \Lambda ,\mathfrak{Z}%
\right) -f_{1}\left( \mathfrak{Z}\right) +U\left( V,\bar{n}\right)
-u^{\prime }\left( \bar{n}\right) \bar{N}.
\end{equation}%
where we have now $\mathfrak{Z}=ze^{\beta \Psi }e^{-\beta u^{\prime }\left(
\bar{n}\right) }$

So, the correction for the Fermi energy in this context is given by:
\begin{equation}
\mu _{0}=\frac{9\pi ^{2}\hbar ^{2}}{8m\mathcal{L}^{2}}\left( \frac{N}{g}%
+1\right) ^{2}-\Psi+u^{\prime }\left(
\bar{n}\right) .
\end{equation}

In contrast to the result derived in  Eq. (\ref{eq:fermi-energy-nonI}), the interaction term appears as an additional term in the above equation. However,  notice that such expression is a generic one. Although we got an analytical result, we cannot proceed  further unless  the structure of the interaction term is explicitly introduced.


\section{Conclusion}

This paper is  aimed at investigating the topological effects on  thermodynamic properties ascribed to a system of moving particles following a closed path through a torus knot. We exploited the grand canonical ensemble framework to study the system of noninteracting particles. Specifically, a fermionic system was studied. From an exact form of the partition function, analytic expressions of thermodynamic observables, e.g.,
Helmholtz free energy, the mean energy, the magnetization and the susceptibility, were obtained and their behavior for a range of parameter values and external conditions was graphically studied. In particular, we discussed in detail how topological features of the torus knot path affected the system.  We also examined  modifications of the Fermi energy. Finally, we outlined a procedure to introduce interaction in the fermionic sector. We hope that some of these characteristic features will be exhibited in near future in a real laboratory experiments, as we have indicated in this work.

\pagebreak
\section*{Acknowledgments}
\hspace{0.5cm}

Particularly, A. A. Araújo Filho acknowledges the Facultad de Física - Universitat de València and Gonzalo J. Olmo for the kind hospitality when part of this work was made. Moreover, this work was partially supported by Conselho Nacional de Desenvolvimento Cient\'{\i}fico e Tecnol\'{o}gico (CNPq) - 142412/2018-0, Coordenação de Aperfeiçoamento de Pessoal de Nível Superior (CAPES) - Finance Code 001, and CAPES-PRINT (PRINT - PROGRAMA INSTITUCIONAL DE INTERNACIONALIZAÇÃO) - 88887.508184/2020-00.


\bibliographystyle{ieeetr}
\bibliography{main}

\end{document}